\documentclass[aip,jmp,preprint,a4paper]{revtex4-1}
\usepackage{cancel}
\usepackage{amsmath}
\usepackage{amssymb}
\usepackage{amsthm}
\usepackage{graphicx}
\usepackage{dcolumn}
\usepackage{bm}
\usepackage{upgreek}
\usepackage{esint}
\usepackage{tikz}
\usetikzlibrary{positioning, fit, arrows.meta, shapes}

\usepackage{commands}

\theoremstyle{definition}
\newtheorem{definition}{Definition}

\theoremstyle{remark}
\newtheorem{remark}{Remark}

\theoremstyle{theorem}
\newtheorem{theorem}{Theorem}

\theoremstyle{lemma}

\theoremstyle{corollary}
\newtheorem{corollary}{Corollary}

\theoremstyle{proposition}

\def\Henon{H\'enonNet }
\def\Henons{H\'enonNets }

\draft 

\begin{document}


\title{Fast neural Poincar\'e maps for toroidal magnetic fields} 



\author{J. W. Burby}
\affiliation{Los Alamos National Laboratory, Los Alamos, New Mexico 87545, USA}
\author{Q. Tang}
\affiliation{Los Alamos National Laboratory, Los Alamos, New Mexico 87545, USA}
\author{R. Maulik}
\affiliation{Argonne National Laboratory, Lemont, Illinois 60439, USA}


\date{\today}

\begin{abstract}
Poincar\'e maps for toroidal magnetic fields are routinely employed to study gross confinement properties in devices built to contain hot plasmas. In most practical applications, evaluating a Poincar\'e map requires numerical integration of a magnetic field line, a process that can be slow and that cannot be easily accelerated using parallel computations. We propose a novel neural network architecture, the H\'enonNet, and show that it is capable of accurately learning realistic Poincar\'e maps from observations of a conventional field-line-following algorithm. After training, such learned Poincar\'e maps evaluate much faster than the field-line integration method. Moreover, the H\'enonNet architecture exactly reproduces the primary physics constraint imposed on field-line Poincar\'e maps: flux preservation. This structure-preserving property is the consequence of each layer in a H\'enonNet being a symplectic map. We demonstrate empirically that a H\'enonNet can learn to mock the confinement properties of a large magnetic island by using coiled hyperbolic invariant manifolds to produce a sticky chaotic region at the desired island location. This suggests a novel approach to designing magnetic fields with good confinement properties that may be more flexible than ensuring confinement using KAM tori.
\end{abstract}


\maketitle 

\section{Introduction}
A commonly used tool for analyzing the behavior of magnetic confinement devices is the magnetic field-line Poincar\'e map.\cite{Abdullaev_2014} Given a toroidal domain $Q$, a poloidal cross section $P\subset Q$, and a magnetic field $\bm{B}$ on $Q$, the Poincar\'e map is constructed by plotting the intersections of magnetic field lines with $P$. The resulting image provides a global summary of the magnetic field line dynamics. Such plots provide a particularly vivid depiction of the intermingling of regular and chaotic field lines, which can be used to rapidly infer gross confinement characteristics of plasma in the device.

The standard computational method for building a Poincar\'e plot involves direct numerical integration of field line trajectories. First a point $\bm{x}$ is chosen in $P$. Then a time-marching algorithm is used to generate an approximate solution of the ordinary differential equation (ODE) $\dot{\bm{x}}(\lambda) =\bm{B}(\bm{x}(\lambda))$ with initial condition $\bm{x}(0) = \bm{x}$. Each time the approximate trajectory crosses $P$, the intersection point is recorded. Because the field line integration could proceed for infinite time, intersections are no longer recorded after some desired number of intersections $N$ is obtained. The whole process is then repeated for different initial conditions in $P$ until a sufficiently-rich Poincar\'e plot has been resolved. { Note that this construction encounters a difficulty if not all field lines intersect $P$ transversally; some magnetic fields, such as those found in reversed field pinches, do not admit a global Poincar\'e section.}

While this process enjoys a high degree of parallelization efficiency due to the decoupling of the integration problems for different initial conditions, an obvious inefficiency is the computational effort spent resolving the field lines between successive intersections with $P$. In order to accurately find intersections, the numerical timestep used for integrating $\dot{\bm{x}}(\lambda) = \bm{B}(\bm{x}(\lambda))$ must be several orders of magnitude smaller than the elapsed time between intersections. Therefore the number of \emph{useful} field line samples generated by the method is several orders of magnitude smaller than the total number of computed samples. In effect, the vast majority of the computation time is spent computing segments of field lines that are not used when generating the Poincar\'e plot.

Theoretically, there is no need to compute long segments of field lines to find a Poincar\'e plot. A basic result from dynamical systems theory states that { if $\bm{B}$ has a nowhere-vanishing toroidal component and $\bm{B}$-lines never leave $Q$} there is a smooth mapping $\Phi:P\rightarrow P$ that sends a point $\bm{x}\in P$ to that point's next intersection with $P$, $\Phi(\bm{x})$.
The mapping $\Phi$ is known as the \emph{first-return map}, or the \emph{Poincar\'e map}. Because $\bm{B}$ is divergence-free, $\Phi$ has the remarkable property that if $U\subset P$ is any region in $P$ then the magnetic flux though $\Phi(U)$ is the same as the magnetic flux through $U$, which says that $\Phi$ is a symplectic mapping relative to a symplectic form\cite{Abraham_2008} on $P$ determined by the magnetic flux. 

Given access to the Poincar\'e map, a Poincar\'e plot could be constructed by iterating $\Phi$ on the same ensemble of initial conditions $\bm{x}\in P$ used in the standard method. Because every iteration of $\Phi$ would generate intersection points on $P$, no amount of computed data would need to be discarded, in contrast to the standard method. While the time required to compute $N$ intersections of a single initial condition using the standard method is $NM\tau_{\text{single-step}}$, where $M$ is the number of timesteps between intersections and $\tau_{\text{single-step}}$ is the time to compute a single integration timestep, the time required to compute the same $N$ intersections using the Poincar\'e map would be $N\tau_{\Phi}$, where $\tau_{\Phi}$ is the time required to evaluate $\Phi$ once. Provided $\tau_\Phi \ll M\tau_{\text{single-step}}$, computing the Poincar\'e plot using the Poincar\'e map would be much faster than the standard approach. Unfortunately, the only general method for computing $\Phi$ available today is equivalent to the standard field-line-following method. (Analytic formulas\cite{Balescu_1998} for $\Phi$ may be found in very special cases.) Obviously, with this method of evaluating $\Phi$, $\tau_{\Phi} = M\tau_{\text{single-step}}$ exactly. 

In this work we will show that deep neural networks can be used to obtain fast, accurate, exactly flux-conserving approximations of Poincar\'e maps $\Phi$ { (when they exist)}  with $\tau_\Phi \ll M\tau_{\text{single-step}}$, where $\tau_{\text{single-step}}$ is the single-step evaluation time for the fourth-order Runge-Kutta (RK4) scheme. Our demonstration will be based on a novel feed-forward network architecture, which we call the H\'enonNet, whose input-to-output mapping is a canonical symplectic map. Like the SympNets introduced in Ref.\,\onlinecite{Jin_2020}, which also feature symplectic input-to-output mappings, a result from Ref.\,\onlinecite{Turaev_2002} implies that H\'enonNets enjoy a symplectic universal approximation property. That is, any symplectic mapping may be approximated by a H\'enonNet arbitrarily well on a given compact set. We find empirically that H\'enonNets are easier to train, and require fewer trainable parameters than SympNets.  We will show that a H\'enonNet may be trained in a supervised fashion by teaching it to reproduce the approximation of a Poincar\'e map $\Phi$ given by field-line integration with the RK4 scheme. After training, the approximation of $\Phi$ provided by the H\'enonNet evaluates orders of magnitude faster than the approximation provided by RK4 integration. Thus, through the use of H\'enonNets, most of the computational overhead associated with field-line integration may be eliminated in a favor of a single off-line training step.

{  While our results specifically target field line flow in magnetic confinement devices, the H\'enonNet architecture may be applied in any setting where canonical symplectic mappings arise. In particular, H\'enonNets may be used as numerical integrators or Poincar\'e-map approximators for any canonical Hamiltonian system.}

{ It is worth emphasizing that the application of a H\'enonNet to field line flow problems generally proceeds through three phases: (1) data-generation, (2) training, (3) prediction. Each phase requires time to complete, and therefore must be accounted for in any cost-benefit analysis that pits H\'enonNets against conventional field-line following routines. Since the training phase (2) is all but guaranteed to be much more time consuming than following a single field line a few times around the torus using a normal field-line integrator, time savings using the H\'enonNet approach should only be expected when a very large number of predictions occur in phase (3). For example, H\'enonNets may dramatically speed up any task where evaluation of the Poincar\'e map occurs as part of the inner-loop of some optimization routine. }

\section{Field-line Poincar\'e maps as canonical symplectic maps}
In order to understand how the H\'enonNets that we will introduce in Section \ref{henon_theory} reproduce the flux-conserving property $\int_{U}\bm{B}\cdot d\bm{S} = \int_{\Phi(U)}\bm{B}\cdot d\bm{S}$ possessed by field-line Poincar\'e maps $\Phi$, it is necessary to understand why flux-preservation is equivalent to the canonical symplectic property. This Section provides a theoretical demonstration of this equivalence. 

Let $Q\subset\mathbb{R}^3$ be a region in $\mathbb{R}^3$ that is diffeomorphic to the solid torus $D^2\times S^1$ with diffeomorphism $(x,y,\phi):Q\rightarrow D^2\times S^1$. (Here $D^2\subset\mathbb{R}^2$ is the standard unit disc and $S^1 = \mathbb{R}/2\pi$ is the $2\pi$-periodic circle.) Assume $\bm{B}$ is a divergence-free field on $Q$ with $\bm{B}\cdot \bm{n}=0$ on $\partial Q$ such that $B^\phi = d\phi(\bm{B})$ is positive. { The latter conditions are typically satisfied in tokamaks and stellarators, at least within the last closed flux surface, but not in reversed field pinches.} {  For $\phi_0\in S^1$, }define $P_{\phi_0} = \{\bm{x}\in Q\mid \phi(\bm{x}) = \phi_0\}$. Because $B^\phi$ is non-vanishing, $P_{\phi_0}$ is a cross section to the $\bm{B}$-flow for each $\phi_0\in S^1$. Therefore there is a well-defined Poincar\'e map, or first-return map $\Phi:P_0\rightarrow P_0$. If $U_0\subset P_0$ ({ $P_0 = P_{\phi_0 = 0}$}) is a compact region in $P_0$ and $T(U_0)$ is the { flux tube connecting $U_0$ with $\Phi(U_0)$}, then $ 0 = \int_{T(U_0)}\nabla\cdot\bm{B}\,d^3\bm{x} = \int_{\Phi(U_0)}\bm{B}\cdot d\bm{S} - \int_{U_0}\bm{B}\cdot d\bm{S} = \Gamma(\Phi(U_0)) - \Gamma(U_0),$ where $\Gamma(U_0)$ is the magnetic flux in the direction of increasing $\phi$ passing through $U_0$. Therefore the Poincar\'e map $\Phi$ preserves magnetic flux. In the following paragraphs, we will use the coordinates $(x,y,\phi)$ to examine this flux conservation property in greater detail. 

If $f$ is any positive smooth function then $\bm{\mathcal{B}} = \bm{B}/f$ has the same streamlines as $\bm{B}$, modulo reparameterization. Therefore $\bm{\mathcal{B}}$ has a Poincar\'e map that is equal to the Poincar\'e map for $\bm{B}$. In particular, we may study the Poincar\'e map $\Phi$ by studying the streamlines of $\bm{\mathcal{B}}$ with $f = B^\phi$.

In the coordinates $(x,y,\phi)$, the vector field $\bm{\mathcal{B}}$ may be written
\begin{align}
\bm{\mathcal{B}} = \mathcal{B}^x\,\partial_x + \mathcal{B}^y\,\partial_y + \partial_\phi,
\end{align}
where $\mathcal{B}^x = B^x/B^\phi,\mathcal{B}^y = B^y/B^\phi$ are smooth functions of $(x,y,\phi)$. Therefore if $(x(\zeta),y(\zeta),\phi(\zeta))$ is a streamline for $\bm{\mathcal{B}}$ then the component functions must satisfy the system of autonomous ordinary differential equations
\begin{align}
\frac{dx}{d\zeta} &= \mathcal{B}^x(x,y,\phi)\\
\frac{dy}{d\zeta}& = \mathcal{B}^y(x,y,\phi)\\
\frac{d\phi}{d\zeta} & = 1.
\end{align}
{ Note that since $\mathcal{\bm{B}} = \bm{B}/B^\phi$ the field-line parameter $\zeta$ differs from the parameter $\lambda$ used in the introduction.} Because $d\phi/d\zeta = 1$, this system of autonomous ODEs on $D^2\times S^1$ is equivalent to the non-autonomous system on $D^2$ given by 
\begin{align}
\frac{dx}{d\phi} & = \mathcal{B}^x_\phi(x,y)\label{na_one}\\
\frac{dy}{d\phi} & = \mathcal{B}^y_\phi(x,y),\label{na_two}
\end{align}
where $\mathcal{B}^i_\phi(x,y) = \mathcal{B}^i(x,y,\phi)$. Note that the ``time" dependence in this non-autonomous system is $2\pi$-periodic. The time-advance map for Eqs.\,\eqref{na_one}-\eqref{na_two} will be denoted $F_{\phi,\phi_0}$; if $(x,y)\in D^2$ then $F_{\phi,\phi_0}(x,y) = (x(\phi),y(\phi))$, where $(x(\phi),y(\phi))$ is the unique solution of Eqs.\,\eqref{na_one}-\eqref{na_two} with $(x(\phi_0),y(\phi_0)) = (x,y)$. Note that the Poincar\'e map may be written in terms of $F_{\phi,\phi_0}$ as $\Phi = F_{2\pi,0}$.

Define the ``time"-dependent $2$-form\cite{MacKay_2020,Bott_1982,Abraham_2008} $\omega_\phi$ on $D^2$ according to 
\begin{align}
\omega_\phi(x,y) = B^\phi(x,y,\phi)\,\mathcal{J}(x,y,\phi)\,dx\wedge dy,
\end{align}
where $\mathcal{J}$ denotes the Jacobian, $d^3\bm{x} = \mathcal{J}\,dx\,dy\,d\phi$. The following argument shows that $\omega_\phi$ is advected by the flow of Eqs.\,\eqref{na_one}-\eqref{na_two}. Because $\bm{B}$ is divergence-free, $\nabla\cdot(B^\phi\bm{\mathcal{B}}) = 0$. In the coordinates $(x,y,\phi)$, the last identity implies
\begin{align}
\partial_\phi(B^\phi\mathcal{J})+\partial_x (B^\phi\,\mathcal{J}\mathcal{B}^x) +\partial_y(B^\phi\mathcal{J}\mathcal{B}^y)  = 0.
\end{align}
If $V_\phi = \mathcal{B}^x_\phi\,\partial_x + \mathcal{B}^y_\phi\,\partial_y$, we therefore have
\begin{align}
\partial_\phi\omega_\phi + L_{V_\phi}\omega_\phi &= \partial_\phi(B^\phi\,\mathcal{J})\,dx\wedge dy + \mathbf{d}\iota_{V_{\phi}}\omega_\phi\nonumber\\
& = \partial_\phi(B^\phi\,\mathcal{J})\,dx\wedge dy + \mathbf{d}\left(\mathcal{B}^x\,B^\phi\,\mathcal{J}\,dy - \mathcal{B}^y\,B^\phi\,\mathcal{J}\,dx\right)\nonumber\\
& = \partial_\phi(B^\phi\,\mathcal{J})\,dx\wedge dy + \partial_x(\mathcal{B}^x\,B^\phi\,\mathcal{J})\,dx\wedge dy - \partial_y(\mathcal{B}^y\,B^\phi\,\mathcal{J})\,dy\wedge dx\nonumber\\
& = \left(\partial_\phi(B^\phi\mathcal{J})+\partial_x (B^\phi\,\mathcal{J}\mathcal{B}^x) +\partial_y(B^\phi\mathcal{J}\mathcal{B})\right)\,dx\wedge dy\nonumber\\
& = 0,
\end{align}
which says that $\omega_\phi$ is advected by $V_\phi$.

The unique solution of the equation $\partial_\phi\omega_\phi + L_{V_\phi}\omega_\phi = 0$ is $\omega_\phi = (F_{\phi,0})_*\omega_0$, where $(F_{\phi,\phi_0})_*$ denotes the pushforward\cite{Abraham_2008} along $F_{\phi,\phi_0}$. Setting $\phi = 2\pi$, we therefore obtain the fundamental result $(F_{2\pi,0})_*\omega_0 = \Phi_*\omega_0 = \omega_0$, or equivalently $\Phi^*\omega_0 = \omega_0$, which says that the mapping $\Phi:D^2\rightarrow D^2$ preserves the $2$-form $\omega_0$. If the component functions of $F_{2\pi,0}$ are denoted $F_{2\pi,0} = (\overline{x},\overline{y})$ then the condition $\Phi^*\omega_0 = \omega_0$ may also be written
\begin{align}
B^\phi(\overline{x},\overline{y},0)\,\mathcal{J}(\overline{x},\overline{y},0)\,d\overline{x}\wedge d\overline{y} = B^\phi(x,y,0)\,\mathcal{J}(x,y,0)\,dx\wedge dy.\label{nc_symplectic}
\end{align}
It is simple to verify that Eq.\,\eqref{nc_symplectic} is the differential form of the integral flux conservation law $\Gamma(U_0) = \Gamma(\Phi(U_0))$ derived earlier. Indeed, by the definition of the surface integral
\begin{align}
\Gamma(U_0) = \int_{U_0}\bm{B}\cdot d\bm{S} = \int_{W_0}\mathcal{J}(x,y,0)\,B^\phi(x,y,0)\,dx\,dy,
\end{align}
where $W_0 $ is the image of $U_0$ in the coordinates $(x,y,\phi)$. Therefore the condition $\Gamma(\Phi(U_0)) = \Gamma(U_0)$ may be written in the form
\begin{align}
 \int_{W_0}\mathcal{J}(x,y,0)\,B^\phi(x,y,0)\,dx\,dy = \int_{\Phi(W_0)}\mathcal{J}(\overline{x},\overline{y},0)\,B^\phi(\overline{x},\overline{y},0)\,d\overline{x}\,d\overline{y}  ,
\end{align}
which implies Eq.\,\eqref{nc_symplectic} because $W_0$ is arbitrary.

We may now give a simple explanation of the sense in which the Poincar\'e map is symplectic. Consider the coordinate transformation $\Psi:(x,y)\mapsto (x,p_x)$, where
\begin{align}
p_x(x,y) = \int_0^y B^\phi(x,\overline{y},0)\,\mathcal{J}(x,\overline{y},0)\,d\overline{y}.
\end{align}
Because $dp_x = \partial_xp_x\,dx + B^\phi(x,y,0)\,\mathcal{J}(x,y,0)\,dy$, the $2$-form $\omega_0 = dx\wedge dp_x$ is canonical with respect to the coordinates $(x,p_x)$. Therefore the condition $\Phi^*\omega_0 = \omega_0$ is equivalent to $\Phi^*(dx\wedge dp_x) = dx\wedge dp_x$, which says that the mapping $\Phi_\Psi = \Psi\circ\Phi\circ\Psi^{-1}$ preserves the canonical symplectic form on $D^2$. In other words, if $\Phi_\Psi(x,p_x) = (\overline{x},\overline{p}_x)$ then $d\overline{x}\wedge d\overline{p}_x = dx\wedge dp_x$.

{ The preservation of the canonical $2$-form $dx\wedge dp_x$ suggests a connection with the theory of Hamiltonian systems. This is no coincidence. Under the assumption $|B^\phi| >0$, the reparameterized field-line dynamics \eqref{na_one}-\eqref{na_two} are equivalent to a $1.5$-degree-of-freedom Hamiltonian system. This well-known fact has been used to analyze field line flow in fusion devices for decades, and could have been used to demonstrate the equivalence between flux conservation and the symplectic property. See Ref.\,\onlinecite{Cary_1983} for a lucid discussion of the connection between field-line flow and Hamiltonian dynamics.}

\section{Henon Networks\label{henon_theory}}
Let $U\subset \mathbb{R}^{n}\times\mathbb{R}^n=\mathbb{R}^{2n}$ be an open set in an even-dimensional Euclidean space. For applications to magnetic fields, $n=1$, but the theory of H\'enonNets applies to all $n$. Denote points in $\mathbb{R}^n\times\mathbb{R}^n$ using the notation $(x,y)$, with $x,y\in\mathbb{R}^n$. A smooth mapping $\Phi:U\rightarrow\mathbb{R}^{2n}$ with components $\Phi(x,y) = (\overline{x}(x,y),\overline{y}(x,y))$ is symplectic if
\begin{align}
\sum_{i=1}^n dx^i\wedge dy^i = \sum_{i=1}^n d\overline{x}^i\wedge d\overline{y}^i.\label{symplectic_property}
\end{align}
The symplectic condition \eqref{symplectic_property} implies that $\Phi$ has a number of special properties. In particular, $\Phi$ must be volume-preserving. (Note, however, that there exist volume-preserving maps that are not symplectic;\cite{Gromov_1985} there is more to the symplectic property than volume preservation when $n\geq 2$.) In spite of the restrictions placed on $\Phi$ by the symplectic condition, the space of all symplectic maps is infinite dimensional.\cite{Weinstein_1971} The problem of finding approximations of an arbitrary symplectic map using compositions of elementary symplectic mappings is therefore inherently interesting.

In Ref.\,\onlinecite{Turaev_2002}, Turaev showed that every symplectic mapping may be approximated arbitrarily well by compositions of \emph{H\'enon maps}, which are special elementary symplectic mappings.
\begin{definition}
Let $V:\mathbb{R}^n\rightarrow\mathbb{R}$ be a smooth function and let $\eta\in\mathbb{R}^n$ be a constant. The \emph{H\'enon map} $H[V,\eta]:\mathbb{R}^n\times\mathbb{R}^n\rightarrow\mathbb{R}^n\times\mathbb{R}^n$ with potential $V$ and shift $\eta$ is given by
\begin{align}
H[V,\eta]\begin{pmatrix} x\\y \end{pmatrix} = \begin{pmatrix} y + \eta\\ -x +\nabla V(y) \end{pmatrix}.
\end{align}
\end{definition}
In particular, we have the following Theorem.
\begin{theorem}[Turaev, 2003]\label{thm1}
Let $\Phi:U\rightarrow\mathbb{R}^n\times\mathbb{R}^n$ be a $C^{r+1}$ symplectic mapping. For each compact $C\subset U$ and $\epsilon >0$ there is a smooth function $V:\mathbb{R}^n\rightarrow\mathbb{R}$, a constant $\eta$, and a positive integer $N$ such that $H[V,\eta]^{4N}$ approximates $\Phi$ within $\epsilon$ in the $C^r$ topology.
\end{theorem}
\begin{remark}
The significance of the number $4$ in this theorem follows from the fact that the fourth iterate of the H\'enon map with trivial potential and shift is the identity map.
\end{remark}

Turaev's result suggests the following method for using deep neural networks to approximate symplectic mappings. First we introduce the notion of a \emph{H\'enon layer}.
\begin{definition}
A \emph{scalar feed-forward neural network} on $\mathbb{R}^n$ is a smooth mapping $V:\mathcal{W}\times\mathbb{R}^n\rightarrow \mathbb{R}$, where $\mathcal{W}$ is a space of network weights. We will use the notation $V[W]$ to denote the mapping $V[W](y) = V(W,y)$, $y\in\mathbb{R}^n$, $W\in\mathcal{W}$.
\end{definition}

\begin{definition}
Let $V$ be a scalar feed-forward neural network on $\mathbb{R}^n$  and let $\eta\in\mathbb{R}^n$ be a constant. The \emph{H\'enon layer} with potential $V$, shift $\eta$, and weight $W$ is the iterated H\'enon map $L[V[W],\eta] = H[V[W],\eta]^4$.
\end{definition}
\noindent There are various network architectures for the potential $V[W]$ that are capable of approximating any smooth function $V:\mathbb{R}^n\rightarrow\mathbb{R}$ with any desired level of accuracy. For example, a fully-connected neural network with a single hidden layer of sufficient { width} can approximate any smooth function. Therefore a corollary of Theorem \ref{thm1} is that any symplectic map may be approximated arbitrarily well by the composition of sufficiently many H\'enon layers with various potentials and shifts. This leads to the notion of a \emph{H\'enon Network}.
\begin{definition}
Let $N$ be a positive integer, let $\bm{V} = \{V_k\}_{k\in \{1,\dots, N\}}$ be a family of scalar feed-forward neural networks on $\mathbb{R}^n$, let $\bm{W} = \{W_k\}_{k\in\{1,\dots,N\}}$ be a family of network weights for $\bm{V}$, and let $\bm{\eta} = \{\eta_k\}_{k\in\{1,\dots,N\}}$ be a family of constants in $\mathbb{R}^n$. The \emph{H\'enon network} with layer potentials $\bm{V}$, layer weights $\bm{W}$, and layer shifts $\bm{\eta}$ is the mapping 
\begin{align}
\mathcal{H}[\bm{V}[\bm{W}],\bm{\eta}] = L[V_N[W_N],\eta_n]\circ\dots\circ L[V_1[W_1],\eta_1].
\end{align}
A graphical summary of the H\'enon network architecture using a single-layer fully-connected architecture for $V(y)$ is given in Fig.\ref{fig:diagram}.
\end{definition}
\begin{remark}
Note that every H\'enon network (H\'enonNet) is a symplectic mapping, regardless of the architectures for the networks $V_k$, and the values of the weights $W_k$. This follows from the simple fact that the composition of symplectic mappings is symplectic. Also note that Turaev's Theorem (Theorem \ref{thm1}) implies the family of H\'enonNets with $V_k = V$, $W_k = W$, $\eta_k = \eta$, $k\in\{1,\dots, N\}$, with $V$ a scalar feed-forward neural network, $W$ a set of weights for $V$, and $\eta\in\mathbb{R}^n$, is sufficiently expressive to approximate any symplectic mapping. However, the number of trainable parameters required to approximate a given symplectic map using such a ``shared-weight" H\'enonNet may not be optimal within the space of all H\'enonNets.
\end{remark}

\begin{corollary}
Let $\Phi:U\rightarrow\mathbb{R}^n\times\mathbb{R}^n$ be a $C^{r+1}$ symplectic mapping. For each compact $C\subset U$ and $\epsilon >0$ there is a H\'enonNet $\mathcal{H}$ that approximates $\Phi$ within $\epsilon$ in the $C^r$ topology.
\end{corollary}

{
\def\layersep{2.5cm}
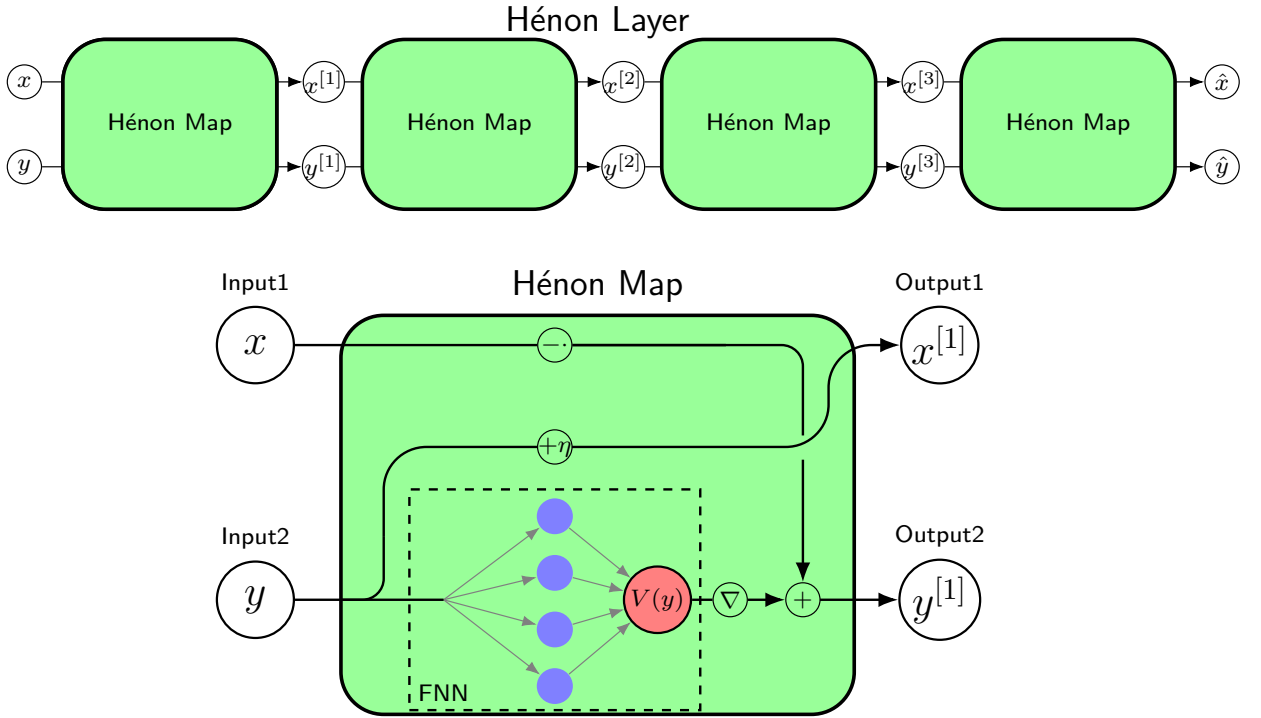
\begin{figure}[htb]
\begin{center}
\resizebox{11.4cm}{!}{
\begin{tikzpicture}[    
    font=\sf \scriptsize,
    >=LaTeX,
    cell/.style={
        rectangle, 
        rounded corners=5mm, 
        draw,
        very thick,
        },
    operator/.style={
        circle,
        draw,
        inner sep=-0.5pt,
        minimum height =.4cm,
        },
    function/.style={
        ellipse,
        draw,
        inner sep=1pt
        },
    ct/.style={
        circle,
        draw,
        line width = .75pt,
        minimum width=.9cm,
        inner sep=1pt,
        },
    gt/.style={
        rectangle,
        draw,
        minimum width=4mm,
        minimum height=3mm,
        inner sep=1pt
        },
    mylabel/.style={
        font=\scriptsize\sffamily
        },
    ArrowC1/.style={
        rounded corners=.25cm,
        thick,
        },
    ArrowC2/.style={
        rounded corners=.5cm,
        thick,
        },]
        
          \useasboundingbox (0,0) rectangle (10,8); 
          
           \begin{scope}[xshift=5cm,yshift=2.4cm]
    \node [cell, fill=green!40, minimum height =4.7cm, minimum width=6cm] at (0,-.5){} ;
     \node [rectangle, dashed, draw,  thick, minimum height =2.6cm, minimum width=3.4cm] at (-0.5,-1.5){} ;
    \node [] at (0.,2.2) {\normalsize H\'enon Map};
     \node [] at (-1.8,-2.6) {FNN};
    
    \node [ circle,draw, line width = .75pt, minimum width=.6cm,inner sep=1pt,fill=red!50] (ibox4) at (0.7,-1.5) {$V(y)$};

    \node [operator] (nId) at (-0.5,1.5) {$-\cdot$};
    \node [operator] (mux2) at (-0.5,0.3) {$+\eta$};
    \node [operator] (mux3) at (2.4,-1.5) {$+$};
    \node [operator] (mux4) at (1.55,-1.5) {$\nabla$};
    
    \coordinate (x) at (-2.5,-0.75);
     \coordinate (y) at (1.5,1.5);
      \coordinate (z) at (2.7,.3);
       \coordinate (fnn) at (-1.8,-1.5);
       \node[] (cross) at (2.4,.3){};

    \node[ct, label={[mylabel]Input1}] (c) at (-4,1.5) {{\large$x$}};
    \node[ct, label={[mylabel]Input2}] (h) at (-4,-1.5) {{\large$y$}};
    

    \node[ct, label={[mylabel]Output1}] (c2) at (4,1.5) {\large${x}^{[1]}$};
    \node[ct, label={[mylabel]Output2}] (h2) at (4,-1.5) {\large${y}^{[1]}$};
    \foreach \name / \y in {1,...,4}
        \node[circle, minimum size=12pt,inner sep=0pt, fill=blue!50] (I-\name) at (-.5, .15-\y/1.5) {};
     
    \foreach \name / \y in {1,...,4}
         \draw [->, thin, draw=black!50] (I-\name)--(ibox4);
         
      \foreach \name / \y in {1,...,4}
          \draw [->, thin, draw=black!50] (fnn)-- (I-\name);
    \draw [ArrowC1] (c) -- (nId) -- (y);
    
    \draw [thick] (h) --(fnn);
    \draw [ArrowC1] (h -| x)++(-0.5,0) -| (x);

    \draw [-, ArrowC2] (x) |- (mux2);
    \draw [->, thick] (ibox4)--(mux4)-- (mux3);
    \draw [-, ArrowC1] (nId) -| (cross);
    \draw [->, thick] (cross)++(0.,-.15) -| (mux3);

    \draw [->, thick] (mux3)++(0.2,0) |- (h2);
    \draw [->, ArrowC2] (mux2)++(0.2,0) --(z) |- (c2);
    \end{scope}

 \node [] at (5.,7.7) {\normalsize H\'enon Layer};
      \begin{scope}[xshift=0cm,yshift=6.5cm]
 \node [operator] (in1) at (-1.7,.5) {$x$};
 \node [operator] (in2) at (-1.7,-.5) {$y$};
 \node [cell, fill=green!40, minimum height =2.cm, minimum width=2.5cm] at (0,0){H\'enon Map} ;

 \node [operator] (in11) at (1.8,.5) {$x^{[1]}$};
 \node [operator] (in12) at (1.8,-.5) {$y^{[1]}$};

 \node [operator] (in21) at (5.3,.5) {$x^{[2]}$};
 \node [operator] (in22) at (5.3,-.5) {$y^{[2]}$};

 \node [operator] (in31) at (8.8,.5) {$x^{[3]}$};
 \node [operator] (in32) at (8.8,-.5) {$y^{[3]}$};
  \node [operator] (ou1) at (12.3,.5) {$\hat x$};
 \node [operator] (ou2) at (12.3,-.5) {$\hat y$};
 
  \draw [->, draw] (in1)-- (in11);
   \draw [->, draw] (in2)-- (in12);
     \draw [->, draw] (in11)-- (in21);
   \draw [->, draw] (in12)-- (in22);
        \draw [->, draw] (in21)-- (in31);
   \draw [->, draw] (in22)-- (in32);
           \draw [->, draw] (in31)-- (ou1);
   \draw [->, draw] (in32)-- (ou2);
   
    \node [cell, fill=green!40, minimum height =2.cm, minimum width=2.5cm] at (0,0){H\'enon Map} ;
     \node [cell, fill=green!40, minimum height =2.cm, minimum width=2.5cm] at (3.5,0){H\'enon Map} ;
     \node [cell, fill=green!40, minimum height =2.cm, minimum width=2.5cm] at (7,0){H\'enon Map} ;
     \node [cell, fill=green!40, minimum height =2.cm, minimum width=2.5cm] at (10.5,0){H\'enon Map} ;

\end{scope}

 %
\end{tikzpicture}
}
\end{center}
\caption{Network diagram. Top: a single H\'enon Layer that consists of four H\'enon maps. Bottom: zoom-in of a single H\'enon map.
\label{fig:diagram}
}
\end{figure}
}

\section{Training setup}
In this work, H\'enon networks are trained to approximate Hamiltonian flows and Poincar\'e maps. 
To generate the training data, a set of points, $\{\mathbf{x}_i\}$, is randomly selected in the domain of interest. 
A well-resolved approximation to Hamiltonian flow maps or Poincar\'e maps is produced using a 4th-order Runge-Kutta method integrating from $t=0$ to $t_0$, generating a set of points,  $\{\mathbf{y}_i\}$. 
For instance, during the training of Poincar\'e maps, we typically use 2000 time steps to integrate the system from 0 to $2\pi$.
This leads to a map from a set of points $\{\mathbf{x}_i\}$ to $\{\mathbf{y}_i\}$.
The mean squared error is then used as the loss function to train H\'enon networks, i.e., 
\begin{align}
MSE= \frac 1 N \sum_{i=1}^N \| \mathcal{H}[\bm{V}[\bm{W}],\bm{\eta}] (\mathbf{x}_i )- \mathbf{y}_i\|^2.
\end{align}
Standard optimization algorithms are then applied to optimize the weights $\bm{W}$ and shifts $\bm{\eta}$. { For the potential networks $V$, we have used fully-connected networks with $\text{tanh}$ activations in each of our examples.}

\noindent\emph{Reproducibility.} Example code used to generate results in this paper is available in Los Alamos technical report LA-UR-20-24873.

We have not explored the use of conventional symplectic integrators for field-line flow to generate the labels $\{\mathbf{y}_i\}$. Provided an RK scheme (or other non-symplectic scheme) is run with sufficient temporal resolution, the benefit of symplectic label generation should be negligible. However, it is plausible that symplectic label generation could offer significant benefits over non-symplectic label generation when the time step is under-resolved. Under-resolved timesteps during the data-generation phase may be desirable as a means of decreasing the overall time for training. 


\section{Approximating Hamiltonian flows using H\'enon Networks\label{ham_flow_sec}}
H\'enon networks have similar structural properties to the SympNets introduced in Ref.\,\onlinecite{Jin_2020}, although the details of the H\'enonNet architecture are substantially different. Before applying H\'enonNets to Poincar\'e maps, it is therefore useful to compare the performance of H\'enonNets with SympNets on a task SympNets have been shown to handle well. This Section presents such a comparison using the implementation of SympNets described in Ref.\,\onlinecite{Jin_2020} and a H\'enonNet whose H\'enon layer potentials $V_k$ are represented as fully-connected neural neworks (FNNs) with a single hidden layer each. The results of this test indicate that the H\'enonNet architecture outperforms the SympNet architecture.

Our test amounts to training both a H\'enonNet and a SympNet to learn the flow map for the mathematical pendulum with fixed timestep parameter $h = 0.1$. The pendulum Hamiltonian is given by 
\begin{align}
H_{\text{p}}(x,y,\phi) = \frac{1}{2}y^2 - \cos x.
\end{align}
Note that the natural (angular) frequency of this pendulum is one, which implies that the timestep $h = 0.1$ is sufficient to fully resolve the pendulum dynamics. Our training setup for each network is the same as in Ref.~\onlinecite{Jin_2020}: 10000 training/test random data points generated in the domain of
$[-\sqrt{2}, \sqrt{2}]\times[-\pi/2, \pi/2]$ with a time step of $h=0.1$.
The performance of the trained models are verified by iterating each approximate flow map 1000 times to
generate the numerical flows of three points in the $(x,y)$ phase space. 
The details of the network architectures used for training are given as follows. The \Henon contains three H\'enon layers, each with its own potential, $V_1,V_2,V_{ 3}$, and shift $\eta_1,\eta_2,\eta_3$. The potentials $V_k(y)$ are parameterized by fully-connected neural networks (FNNs) with one hidden layer each, and $\tanh$ activation. The hidden layers that specify the $V_k$ each have $5$ neurons. This gives a total of $3\times16=48$ trainable parameters. 
On the other hand, the SympNet has a network structure of 
\[
\Phi = \mathcal{L}_n^{(k+1)}\circ (\mathcal{N}_{\text{up/low}}\circ\mathcal{L}_n^{(k)})\circ\dots(\mathcal{N}_{\text{up/low}}\circ\mathcal{L}_n^{(1)}),
\]
where, as described in detail in Ref.\onlinecite{Jin_2020}, each $\mathcal{L}_n^{(k)}$ is the composition of $n$ trainable linear symplectic layers, and $\mathcal{N}_{\text{up/low}}$ is a non-trainable symplectic activation map.
In this work, we use $k=8$ and $n=6$, which corresponds to 8 layers with 6 sub-layers following the definition in Ref.~\onlinecite{Jin_2020}.
Since all the trainable parameters in SympNets are from the linear layers, this gives a total of $9\times8=72$ trainable parameters.

The other hyperparameters for our learning task are given as follows. Our training runs for each network comprise 5000 epochs using the Adam optimizer with a batch size of 1000. 
The learning rate is a piecewise constant-decaying function with the initial rate of $0.1$. We note that in the training of Ref.~\onlinecite{Jin_2020}, $10^6\gg 5000$ epochs were used to produce a good numerical result, which has been confirmed in our own implementation of SympNets. 
Here, however, we intentionally reduce the total number of epochs to examine the performance of the two networks when subject to more modest training regimens.

The test results are shown in Figure~\ref{fig:HamilPendulum}. 
As indicated by the trajectories of three testing points, we note that both networks produce an approximation to the original Hamiltonian that preserves the total energy well. 
However, the accuracy of the H\'enonNet is much better than the accuracy of the SympNet. This is consistent with the final losses of two trainings. 
We note that the training loss for the H\'enonNet is 2.8432e-07 and the test loss is 2.7853e-07, while the losses for the SympNet are 1.9487e-05 and 1.9629e-05, respectively. 
To achieve comparable losses to the H\'enonNet, the SympNet needs about $10^6$ epochs. 
The histories of losses are also presented in Figure~\ref{fig:HamilPendulum}. 
We note that the \Henon loss decays much faster than that of the SympNet, despite the fact that the SympNet has more trainable parameters than the \Henon in this test.
All the observations indicate that  the training of \Henons is easier than the training of SympNets.

To have a fair comparison, we consider the total training time of two models. All the networks described in this work are 
implemented using Tensorflow v2 and run on a single Nvidia V100 (`Volta') GPU for a better performance. 
Due to the more complicated network structure of H\'enonNets relative to SympNets, the training of the H\'enonNet (366s) requires more time than training the SympNet (236s). 
However, considering the significant improvement of the losses,
\Henons still appear to be a much more efficient method to achieve a given MSE loss.




{
\begin{figure}[htp]
\begin{center}
\includegraphics[width=.42\textwidth]{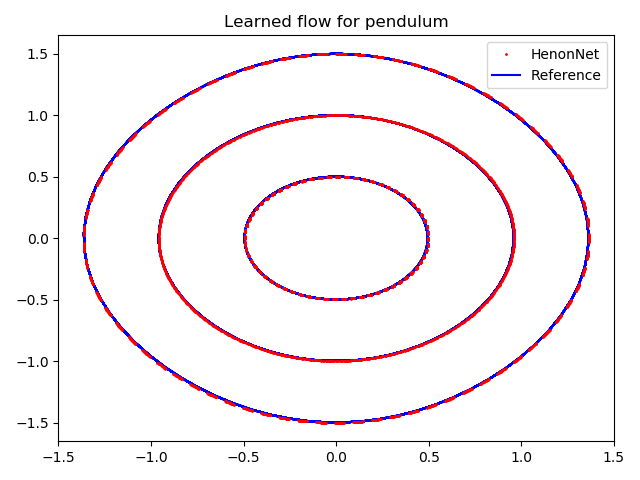}
\hspace{.05\textwidth}
\includegraphics[width=.42\textwidth]{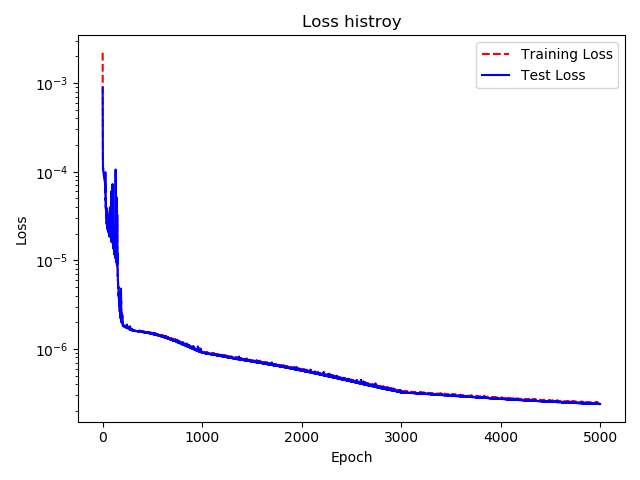}
\includegraphics[width=.42\textwidth]{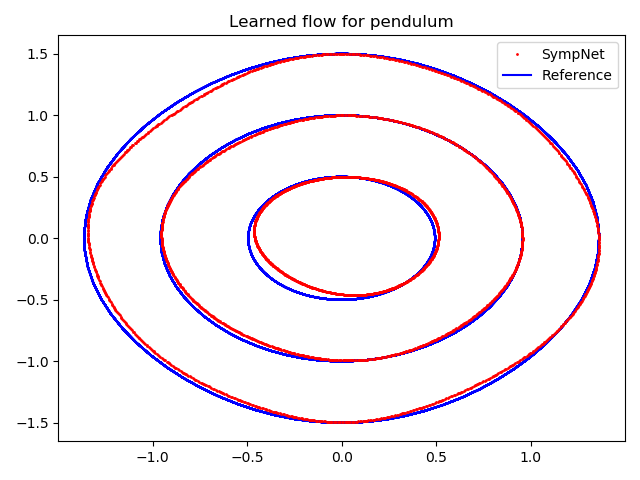}
\hspace{.05\textwidth}
\includegraphics[width=.42\textwidth]{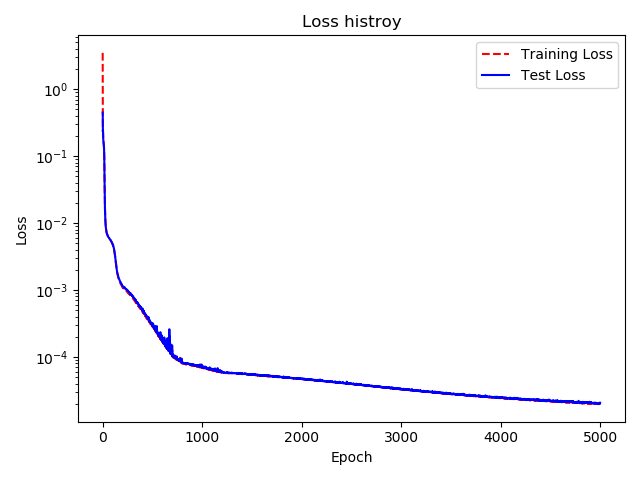}
\end{center}
\caption{Numerical flows and training/test loss histories of \Henon (top row) and SympNet (bottom row) for pendulum. 
The reference is a high-order numerical solution generated by RK4.
\label{fig:HamilPendulum}
}
\end{figure}
}

To validate the aforementioned conclusions, we perform a systematic hyperparameter study for both the \Henon and the SympNet architectures with regards to validation losses. The purpose of this study is to show the superiority of one architecture over the other for an arbitrary hyperparameter choice within a high-dimensional space. To that end, we define a multidimensional space of hyperparameters from which a novel architecture can be randomly chosen by sampling without replacement. This space comprises choices for the batch size, the number of training epochs, the number of layers in an architecture and the number of neurons in each layer of the architecture. We eschew the use of Bayesian optimization to obtain the \emph{best} possible \Henon or SympNet architecture in favor of comparing the general performance of the two for an arbitrary architecture selection. The range of choices for the different hyperparameters are given by
\begin{flalign*}
    &\text{Batch size}: [600,1000,1400,2000], \\
    &\text{Training epochs}: [3500,4000,4500,5500,6000], \\
    &\text{Number of neurons}: [2,5,8,11,14,17], \\
    &\text{Number of layers}: [2,3,4,5,6,8], 
\end{flalign*}
which are common to both the SympNets and the H\'enonNets. Training and validation assessments were also carried out on a single Nvidia V100 GPU with hyperparameters evaluated sequentially for 6 hours each for both types of architectures. While this compute expense was insufficient for assessing all the possible hyperparameter combinations in the high-dimensional space (720), we performed hyperparameter pruning everytime 5 architecture evaluations were completed. This pruning was performed by analyzing the bottom 20\% of the poorest performing architectures (in our case 1 out of 5) and identifying similar architectures that were yet to be sampled. These architectures were then removed from the search space based on a Spearman rank order correlation. The results for the validation loss of the different \Henons and SympNets are shown in Figure \ref{fig:HamilPendulumHPS}. We plot the distribution of the common logarithm of the validation loss for all our evaluated architectures and it is clearly observed that the \Henons outperform the SympNets. In addition, we also provide box-plots for both types of evaluated architectures. The blue line within the box-plot is the median value for validation loss, the green box indicates the interquartile range of the sampled architectures and the whiskers indicate the outliers. These box-plots also show that the \Henons are superior. In particular, the median value of SympNet performance is seen to be beyond the upper whisker of the \Henons outlier range thereby reinforcing our conclusions.

A firm theoretical explanation for the apparent performance gap between H\'enonNets and SympNets is currently lacking. Therefore our empirical observations need to be taken with a grain of salt. That said, there is a fundamental architectural difference between H\'enonNets and SympNets that likely has an important effect on network performance: the per-layer expressive power. A single layer in a SympNet comprises a composition of several trainable linear symplectic maps followed by a single, non-trainable symplectic nonlinear activation. The dimension of the space of maps that may be approximated by a single layer in a SympNet is therefore \emph{at most} $n(2n+1)$, i.e. the dimension of the linear symplectic group $\text{Sp}(2n,\mathbb{R})$, where $n$ is the number of degrees of freedom. In contrast, since a H\'enon map is parameterized by a single free-function $V\in C^\infty(\mathbb{R})$ (along with a bias $\eta\in\mathbb{R}^n$), a single H\'enon layer is capable of representing spaces of symplectic mappings of arbitrarily large dimension. Therefore a single H\'enon layer can provide much more expressive power than a single SympNet layer. In light of these observations, we conjecture that a SympNet requires significantly more layer depth in general than a H\'enonNet to approximate a given symplectic mapping within a specified error tolerance. This conjecture may play a role in explaining the observed performance gap.

{
\begin{figure}[htp]
\begin{center}
\includegraphics[width=.48\textwidth]{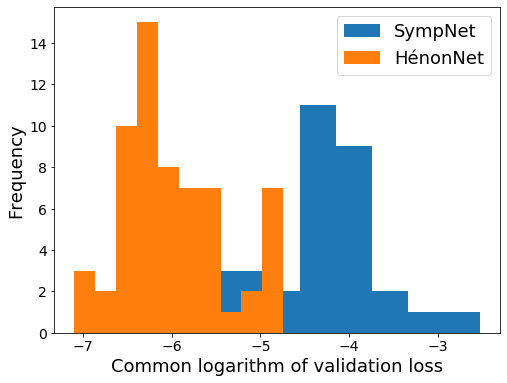}
\includegraphics[width=.48\textwidth]{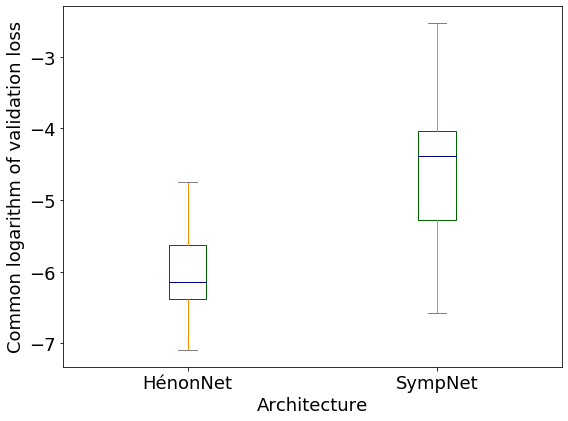}
\end{center}
\caption{Distributions for validation losses for evaluated hyperparameters of both \Henon and SympNet formulations (left) and corresponding box-plots (right) showing median (blue line within box), quartiles (limits of the green box) and outlier (whiskers) ranges. The distributions and box-plots indicate that the average performance of the \Henon is superior to the SympNet within the chosen hyperparameter range. 
\label{fig:HamilPendulumHPS}
}
\end{figure}
}

\section{Poincar\'e maps using H\'enon Networks}
In this section, we consider approximations to Poincar\'e maps using H\'enonNets. Using neural networks to approximate Poincar\'e maps is more ambitious than using networks to approximate a small-timestep Hamiltonian flow. The first challenge is that a larger timestep is used to generate the Poincar\'e map,
which indicates a much stiffer problem than the small-timestep case. For instance, a large time step of $2\pi$ is used to generate the training data in this section, while
in the previous case, the time step is $0.1$. 
An alternative way to view the challenge is that a Poincar\'e map tends to be further away from the identity map than a small-timestep Hamiltonian flow map.  
The second challenge arises in conjunction with the deterministic chaos found in 1.5-degree-of-freedom Hamiltonian systems.
Chaotic regions are characterized by positive Lyapunov exponents $\lambda$. Therefore the separation of nearby trajectories as a function of timestep is proportional to $\exp(h\lambda)$. It is much more likely that a Poincar\'e map experiences an e-folding ($\sim \exp(2\pi\lambda)$) than a small timestep flow map ($\sim \exp(0.1\lambda)$). 
Both of these challenges make training much more difficult than in the test described in Section \ref{ham_flow_sec}.


{ In the remainder of this Section, we will describe results from several numerical experiments designed to explore the capabilities of H\'enonNets. We remark that the hyperparameters chosen for these test H\'enonNets have not been optimized to maximize the quality of our training outcomes. Future applications of H\'enonNets may benefit from further hyperparameter tuning.}

\subsection{Pendulum}
In order to assess the ability of \Henons to approximate Poincar\'e maps for integrable magnetic fields with separatrices, we first considered the pendulum Hamiltonian, 
\begin{align}
H_{\text{p}}(x,y,\phi) = \frac{1}{2}y^2 - \cos x.
\end{align}
Note that the natural angular frequency of the pendulum is $1$. For simplicity, we focus on using  \Henons to approximate the Poincar\'e map inside the separatrix
of the pendulum. 

Here a total of $200$K training points are generated in the domain of $r\le1.5$. 
The training points are split into two groups. The first group consists of $100$K random points inside the region of $r\le0.3$, while the rest of the training points are randomly drawn
in the entire region. 
The reason to cluster points around the origin is that we found it is harder to achieve an accurate approximation to the Poincar\'e  map for the pendulum around the origin.
This is due to the fact the inner points travel a much smaller distance than the outer points (the map around inner points is close to the identity), which means that the near-origin data samples have a disproportionately small contribution to the mean-squared error. 
 Therefore, a small perturbation 
around the origin can lead to the Poincar\'e plot of the pendulum being distorted quite significantly. 
We therefore add more points around the origin to increase the accuracy there. An alternative to this technique, which we have not explored, would be to modify the cost function by increasing the weight of near-origin contributions. We anticipate that this difficulty, as well as our proposed coping mechanism, will be relevant to learning Poincar\'e maps for any magnetic field whose rotational transform tends to zero on-axis.
The second change is that the training region is chosen to be a disk. 
If a box training region is used, the optimization algorithm will focus unnecessary attention to the corners in order to achieve a good approximation
around them (corners are typically highly distorted), which is by no means the goal of the current work as the region of interest is typically far away from the boundary. 
Those two strategies, which appear to be quite general,  accelerate the training significantly. 
Using those training points, the training labels are then generated by a RK4 time integrator with a time interval of $2\pi$ and 2000 intermediate time steps. This forms a Poincar\'e map from the training points to the corresponding locations at $h=2\pi$. 

The \Henon in this case consists of 10 H\'enon layers, each of which has a single-hidden-layer FNN potential with 10 neurons. 
This gives a total of 310 trainable parameters. 
The H\'enonNets with many layers are found to be necessary to achieve a good approximation to the  Poincar\'e map. { This observation is consistent with the fact that a H\'enonNet may be viewed as a ResNet\cite{He_resnet_2015} with additional structure; ResNets are known to achieve higher performance with more layer depth. } 
We use the Adam optimizer with 20000 epochs and a decaying learning rate. The initial learning rate is set to be 0.02, and the batch size is 1000. The training finishes with a final training loss of 2.0776e-07. { We computed a test loss using $2e05$ uniform samples of the disk region $\sqrt{x^2 + y^2}\leq 1.5$ and find a value of 4.0123e-07. If we sample the region near the axis more heavily, as in the training data, the error becomes 2.0657e-07.  }
The trained model is then verified through generating a Poincar\'e plot starting from 20 points along $x$-axis.
The model is used to predict 2000 times recursively, mimicking the process of generating a  Poincar\'e plot using a conventional time integrator.

The results are presented in Figure~\ref{fig:PoincarePendulum}. 
The plot on the left features the Poincar\'e plot generated by the \Henon as well as its starting points (red dots). 
The plot in the middle features the reference Poincar\'e plot generated by RK4.
Two plots are compared to each other on the right.  
We note that the  Poincar\'e plot by the \Henon matches very well with the reference except for the region close to the origin.
For each trajectory, the \Henon preserves the total energy very well, while, on the other hand, the RK integrator requires  a much smaller time step to preserve the same level of total energy (about 100 time steps for each Poincar\'e map). 
Figure~\ref{fig:PoincarePendulum2} presents a typical trajectory in the Poincar\'e plot and its corresponding Hamiltonian.
The errors in Hamiltonian are comparable between the learned model and the RK result.
Note that here the dots stands for time series in the Poincar\'e plot. The time between each dot is $2\pi$.

As a final check, we compare the computational time to produce the Poincar\'e plot during the online stage. 
To have a fair comparison, we use a CPU node to perform the prediction of the \Henon and the RK4 time stepping. 
Both of the algorithms are implemented in Python and the RK solver is vectorized (vectorized operations in Python means arrays operations are performed using optimized, compiled C code).
In the RK case, 100 time steps are used to evolve each Poincar\'e map.
 The \Henon needs 2.6 seconds to finish the prediction of 2000 iterations
while the vectorized RK integrator needs 7.8 seconds. Note that the right hand side of the ODE in this case is very simple, and for more complicated right hand sides,
the speedup will be increased since the \Henon will approximate the right hand side while the RK integrator needs to evaluate it in each time step. 



{
\begin{figure}[htp]
\begin{center}
\includegraphics[width=.32\textwidth]{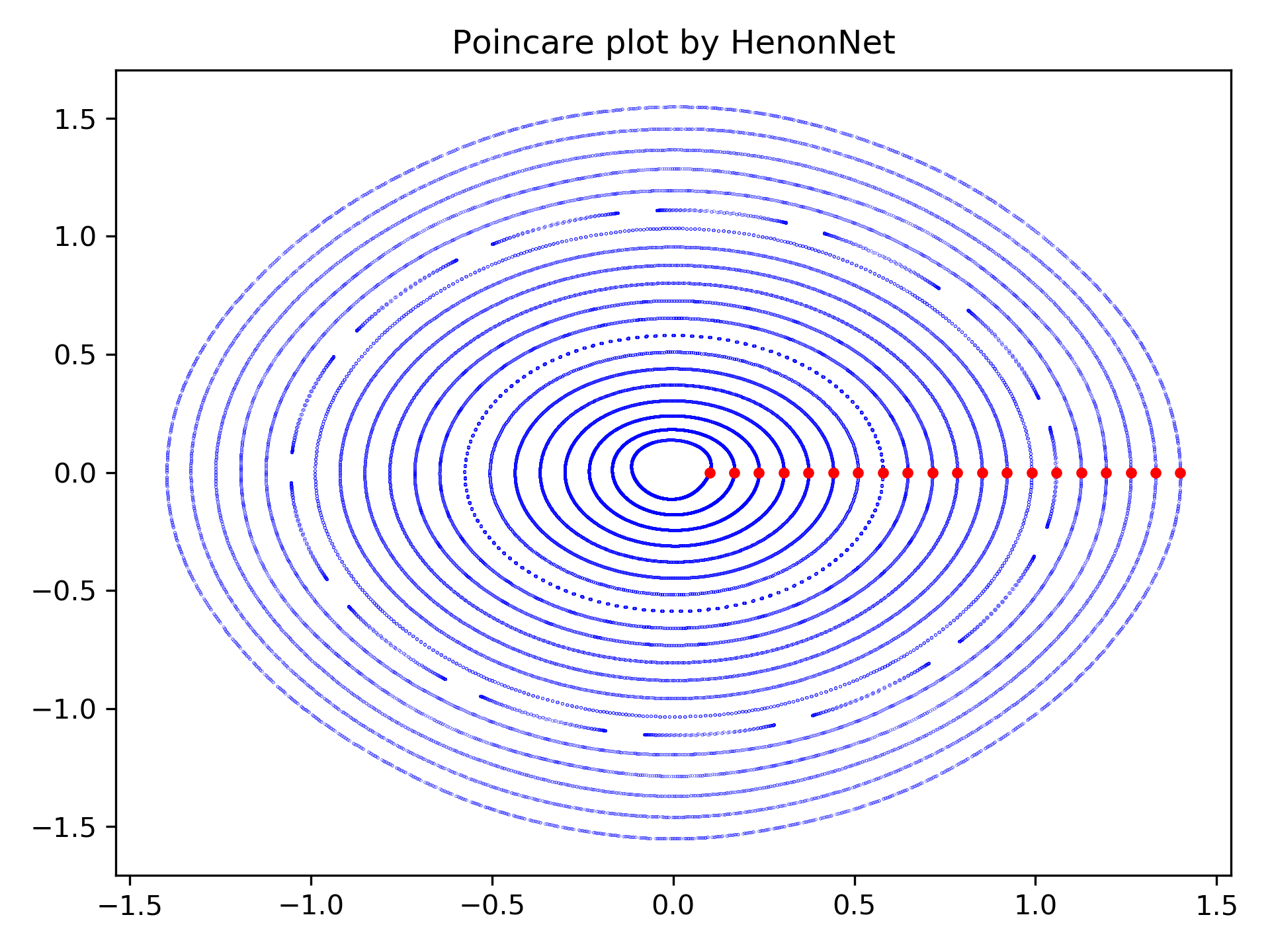}
\includegraphics[width=.32\textwidth]{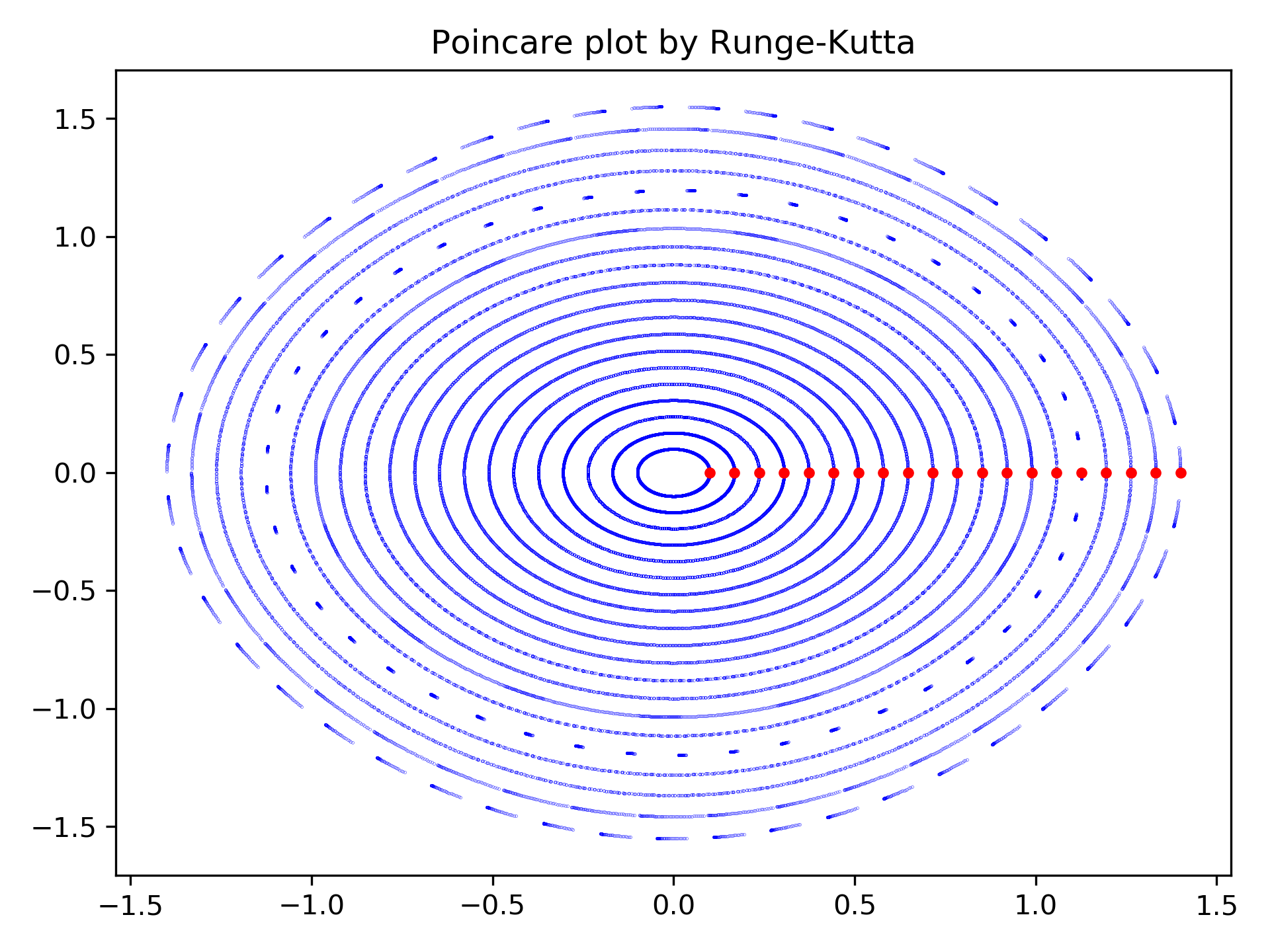}
\includegraphics[width=.32\textwidth]{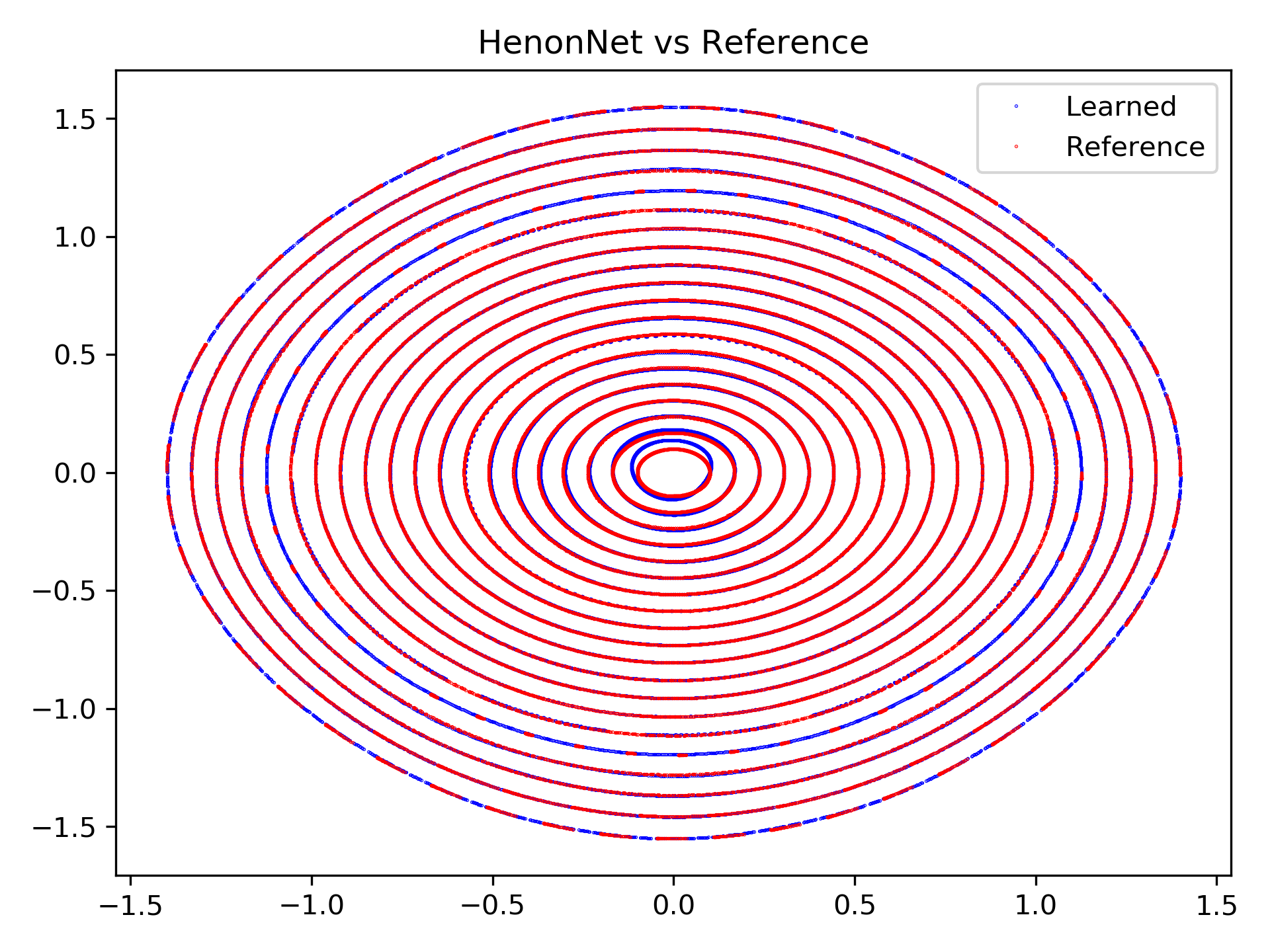}
\end{center}
\caption{Poincar\'e plots of pendulum. Left: Poincar\'e plot generated by the trained H\'enonNet. Middle: Poincar\'e plot generated by RK4. Right: comparing two Poincar\'e plots.
\label{fig:PoincarePendulum}
}
\end{figure}
}

{
\begin{figure}[htp]
\begin{center}
\includegraphics[width=.8\textwidth]{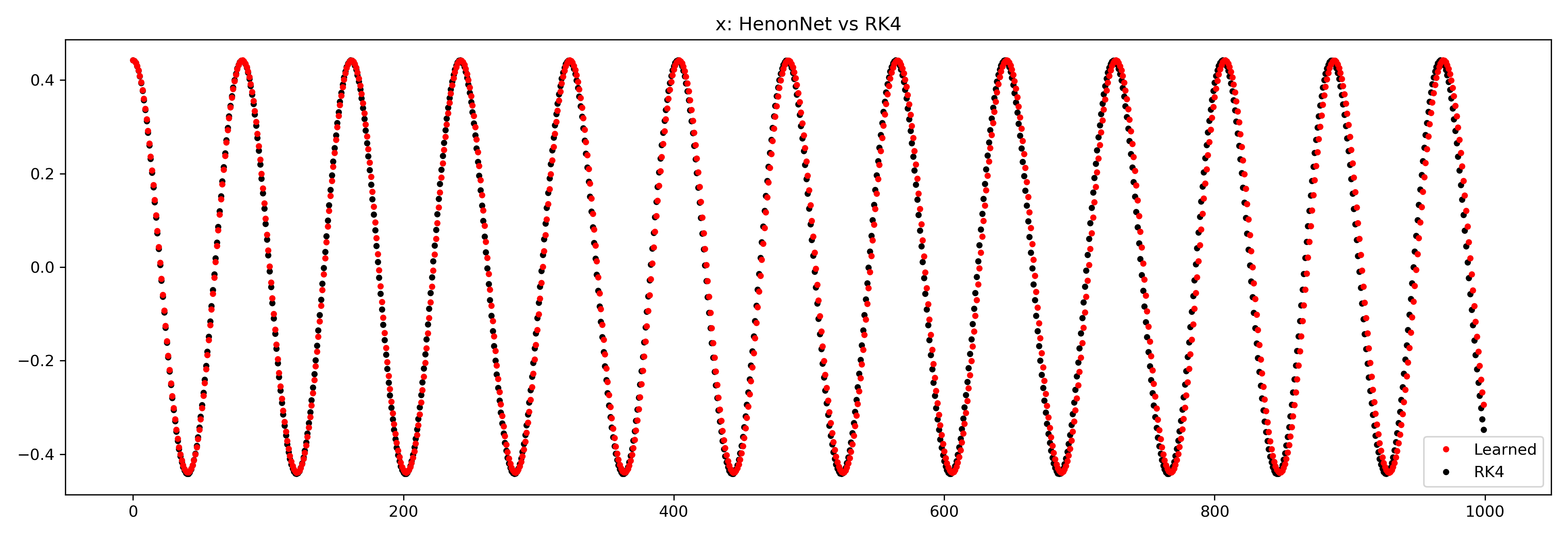}\\
\includegraphics[width=.8\textwidth]{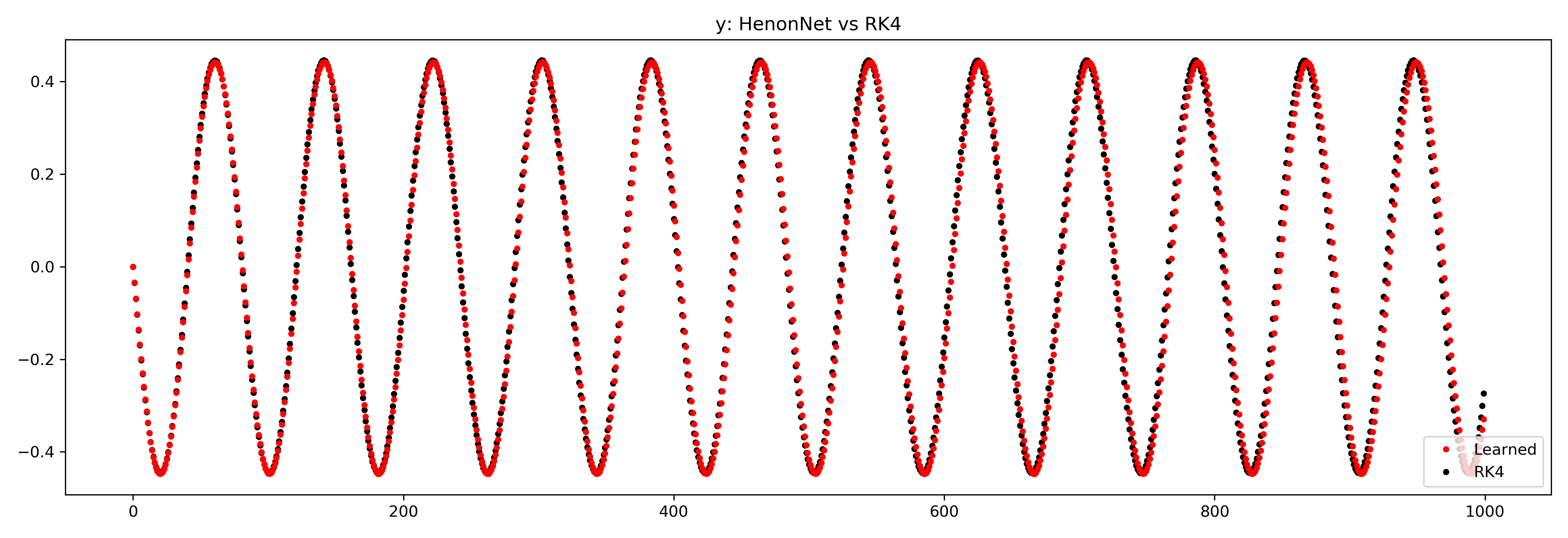}\\
\includegraphics[width=.8\textwidth]{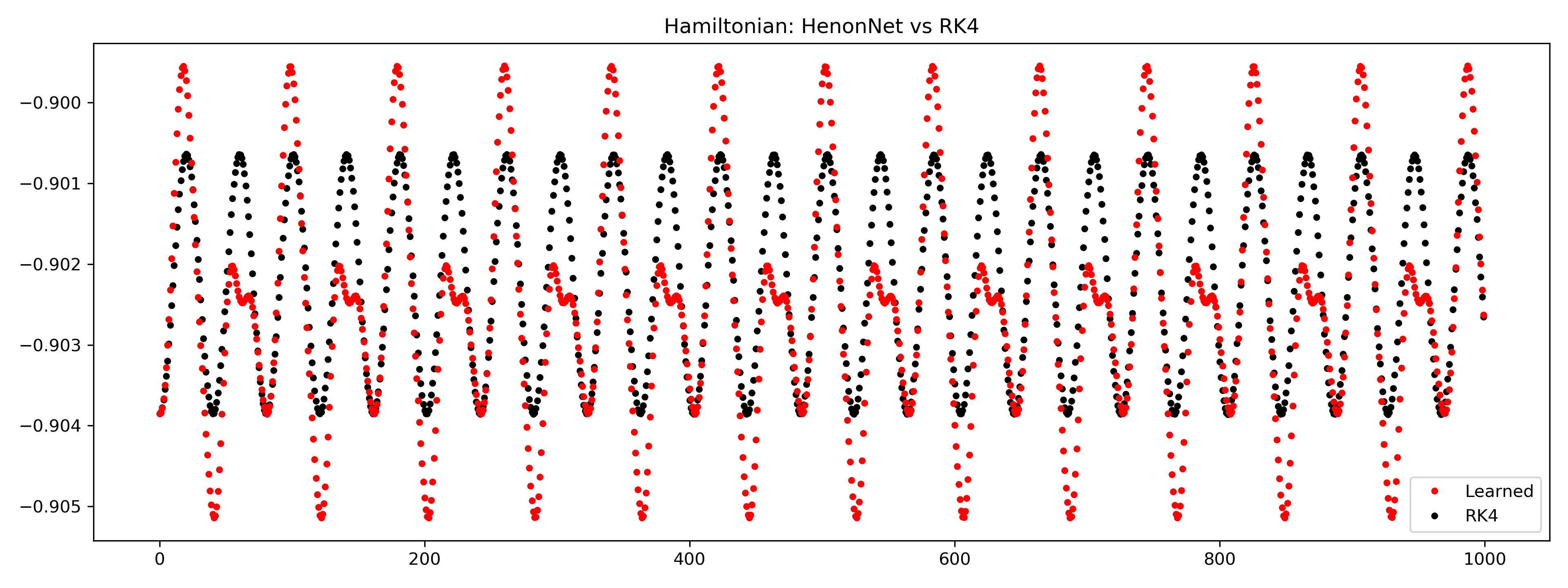}
\end{center}
\caption{Learned $(x, y)$  and Hamiltonian of pendulum. Top: learned $x$ vs { RK4}. Middle: learned $y$ vs { RK4}. Bottom: learned Hamiltonian vs { RK4}. { Horizontal axis is number of iterations of the Poincar\'e map.}
 Here the starting point is $(0.4421, 0)$.
 The dots here stand for the time series in the Poincar\'e plot. { Energy error in the RK4 simulation is a finite-timestep effect. }
\label{fig:PoincarePendulum2}
}
\end{figure}
}

\subsection{Perturbed pendulum}
In order to study the ability of H\'enonNets to approximate the Poincar\'e maps of more complicated magnetic fields with regular phase portraits, we considered the Hamiltonian 
\begin{align}
H_{\text{pp}}(x,y,\phi) = \frac{1}{2}y^2 - \omega_0^2\cos x  -  \epsilon \, \bigg[0.3\,xy \sin(2 \phi) + 0.7\,xy \sin(3 \phi)\bigg],
\end{align}
which represents a perturbed pendulum with natural frequency $\omega_0$. In the test, we choose $\omega_0=0.5$ and $\epsilon=0.5$. Numerical results from highly-resolved Runge-Kutta integration suggest that this system is integrable. However, we cannot prove this because we have failed to identify a first integral.

Here a total of $220$K training points are randomly selected in a disk of $r\le0.9$. 
A Poincar\'e map is generated by a well-resolved RK4 approximation. 
The \Henon in this case consists of 10 H\'enon layers, each of which has 10 neurons in its single-hidden-layer FNN layer potential. 
We use the Adam optimizer with 5000 epochs and a decaying learning rate. The initial learning rate is set to be 0.05, and the batch size is 1000. The training finishes with a final loss of  1.6216e-7. { We evaluated the loss on a set of 3e05 test samples uniformly sampled from the disk $\sqrt{x^2+y^2}\leq 0.9$ and found a value of 1.3978e-07.} Compared to the previous case, this case turns out to be much easier to train. The reason is the rotational transform on-axis is well away from zero, which makes
 the resulting Poincar\'e map far away from
the identity for any given points in the domain.

The trained model is then verified through generating a Poincar\'e plot starting from 20 points along $x$-axis and 10 points along $y$-axis.
The model is used to predict 1000 times recursively, mimicking the process of generating a  Poincar\'e plot using a conventional time integrator.
The results are presented in Figure~\ref{fig:perturbedPendulum}. 
The plot on the left features the Poincar\'e plot generated by the \Henon as well as its starting points (red dots). 
The plot in the middle features the reference Poincar\'e plot generated by RK4. 
Two plots are compared to each other on the right.  
We note that two plots match extremely well with each other, in spite of the presence of interesting islands structures.

We again compare the computational time to produce the Poincar\'e plot. 
 The \Henon needs 1.3 seconds to finish the prediction of 1000 iterations
while the vectorized RK integrator needs 12.1 seconds with 100 time steps for each Poincar\'e map. 
Note that the \Henon has the same structure as the previous case, and it is found that the computational time is consistent with the previous case (it uses about half of the previous time
since the iterations are half of the previous case).
On the other hand, the time of the RK integrator grows due to the growing complexity of the right hand side.

{
\begin{figure}[htp]
\begin{center}
\includegraphics[width=.32\textwidth]{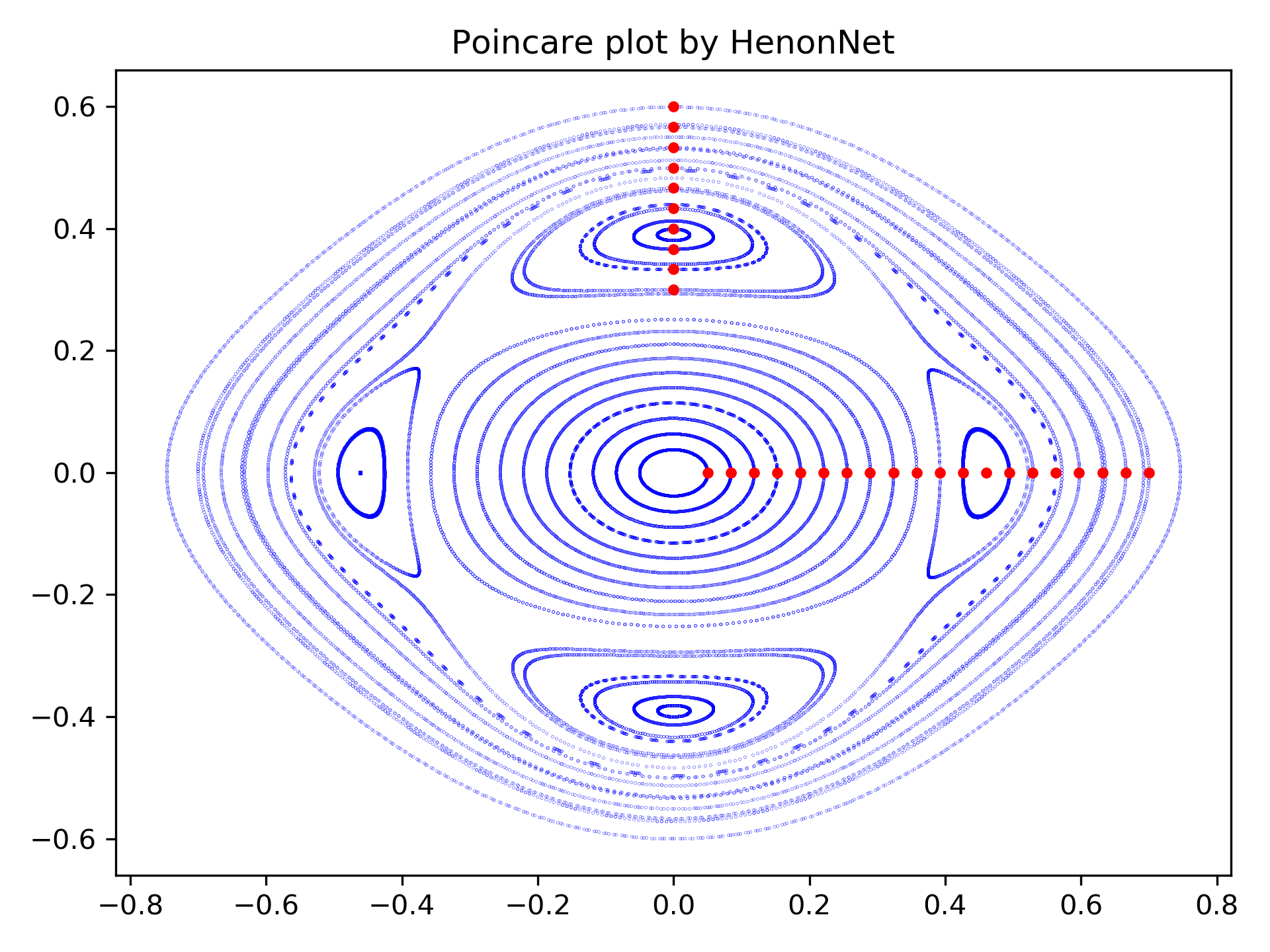}
\includegraphics[width=.32\textwidth]{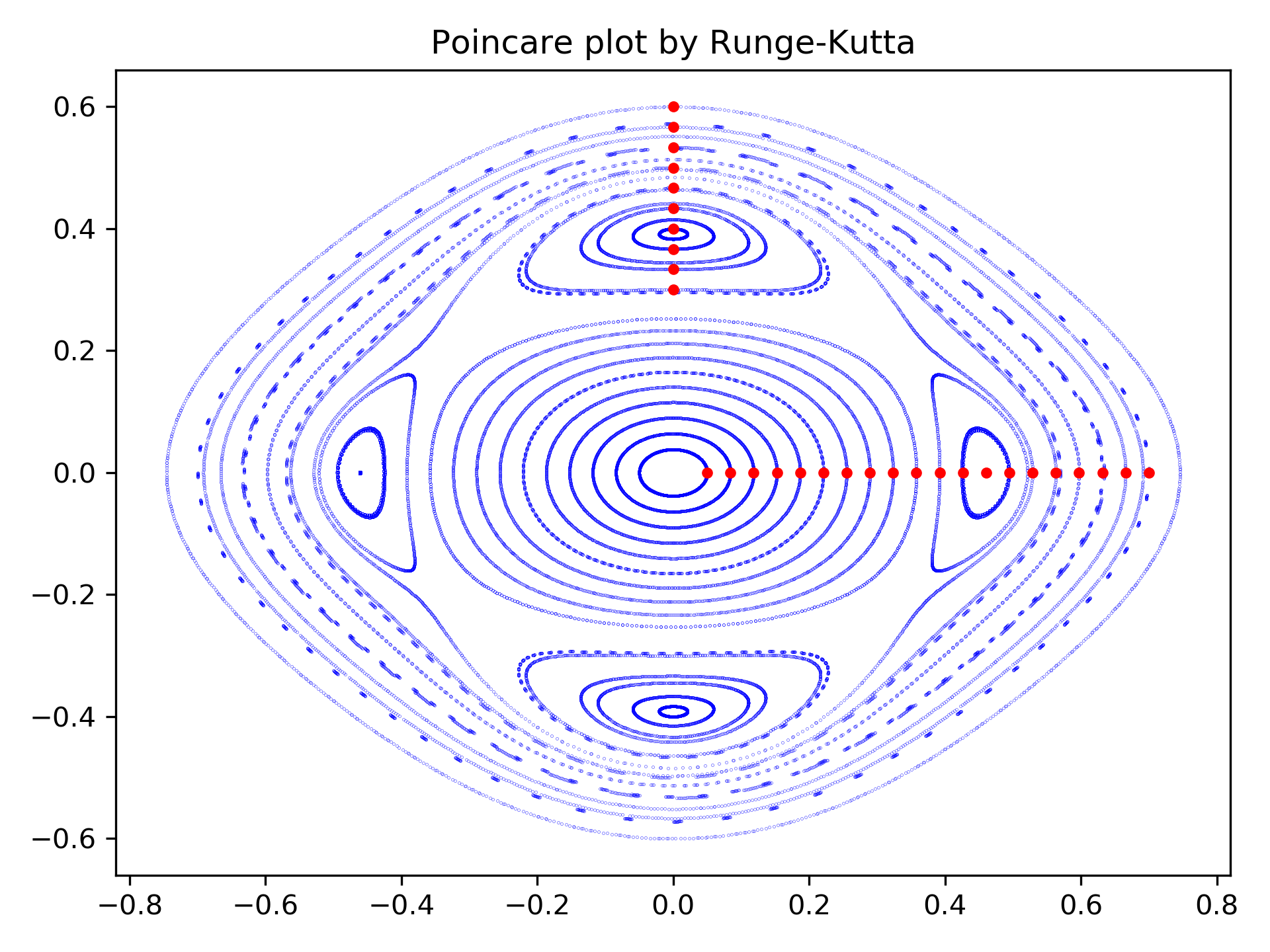}
\includegraphics[width=.32\textwidth]{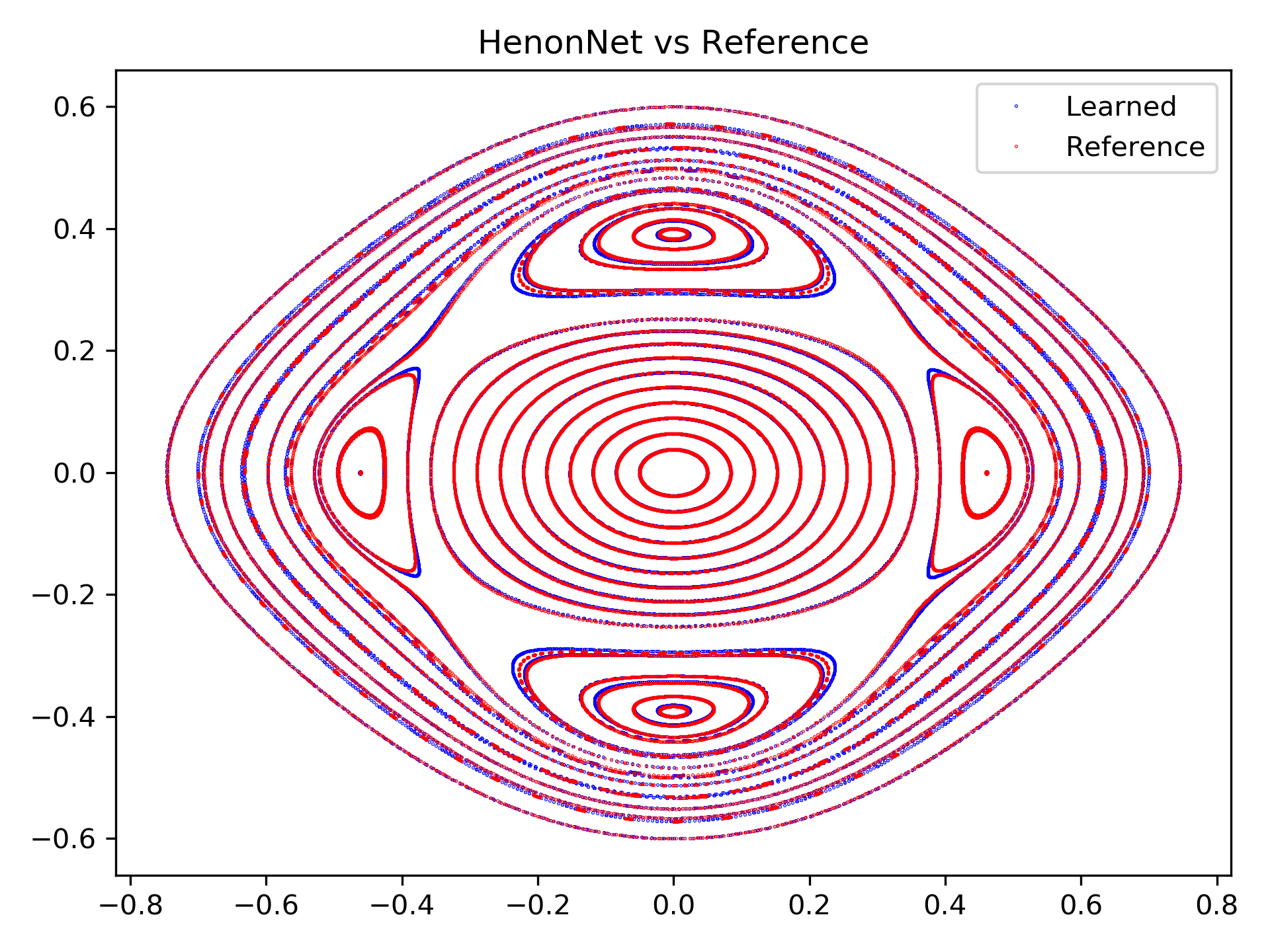}
\end{center}
\caption{Poincar\'e plots of perturbed pendulum. Left: Poincar\'e plot generated by the trained H\'enonNet. Middle: Poincar\'e plot generated by RK4. Right: comparing two Poincar\'e plots.
\label{fig:perturbedPendulum}
}
\end{figure}
}



\subsection{Prototypical resonant magnetic perturbation}
As a final test, we demonstrate the ability of \Henons to handle realistic  magnetic fields in magnetic confinement devices. We consider a non-integrable Hamilonian that mimics magnetic fields subject to large resonant magnetic perturbations (RMP).  { The unpertubed Hamiltonian corresponds to field lines with a fixed irrational rotational transform. The toroidal-angle-average of the perturbing Hamiltonian produces a small amount of shear that leads to the creation of many rational surfaces. These rational surfaces then resonate with the fluctuating part of the perturbing Hamiltonian, thereby mimicking the effect of externally-applied error fields in modern tokamak experiments.}
{ More precisely,} the Hamiltonian is given by
\begin{align}
H_{\text{RMP}}(x,y,\phi) &= \frac{1}{2}y^2 + \frac{1}{4}x^2\nonumber\\
& + \frac{\epsilon}{4}\bigg(\tanh\left((x-y)x^2\cos(3\phi)\right) + \frac{1}{5}\tanh\left((x-y)x^2\sin(3\phi)\right)\bigg),\label{HRMP}
\end{align}
which represents a Harmonic oscillator with natural angular frequency $1/\sqrt{2}$ subject to a resonant, time-dependent perturbation.
Here the perturbation amplitude is set to be $\epsilon=0.25$.

A total of $400$K sampling points are selected to generate the training data. As in the previous examples, labels are assigned to samples using a well-resolved RK4 approximation of the Poincar\'e map.
We use a simple strategy  to select the data points such that the resulting training data covers the domain of interest and meanwhile minimizes 
the unnecessary points outside of the  domain.
First, 10K training points are randomly selected in an ellipse of radii  1.75 and 1. Then the 10K points are used as the seed to generate its Poincar\'e map recursively 400 times. 
Every ten iterations we collect the corresponding Poincar\'e map generated at the current iteration, which contributes to 400K training points in total. 
The input and output points of the Poincar\'e map, used as the training data, are presented in Figure~\ref{fig:rmp-data}.
Note that due to the selection strategy the training data starts to form a similar pattern as the Poincar\'e plot.
This strategy can be viewed as a generalization of the previous case where a disk training region is used. 
It is found such a strategy significantly improves the efficiency of training relative to choosing samples from a region that is not approximately invariant under the discrete-time flow. 

{
\begin{figure}[htp]
\begin{center}
\includegraphics[width=.4\textwidth]{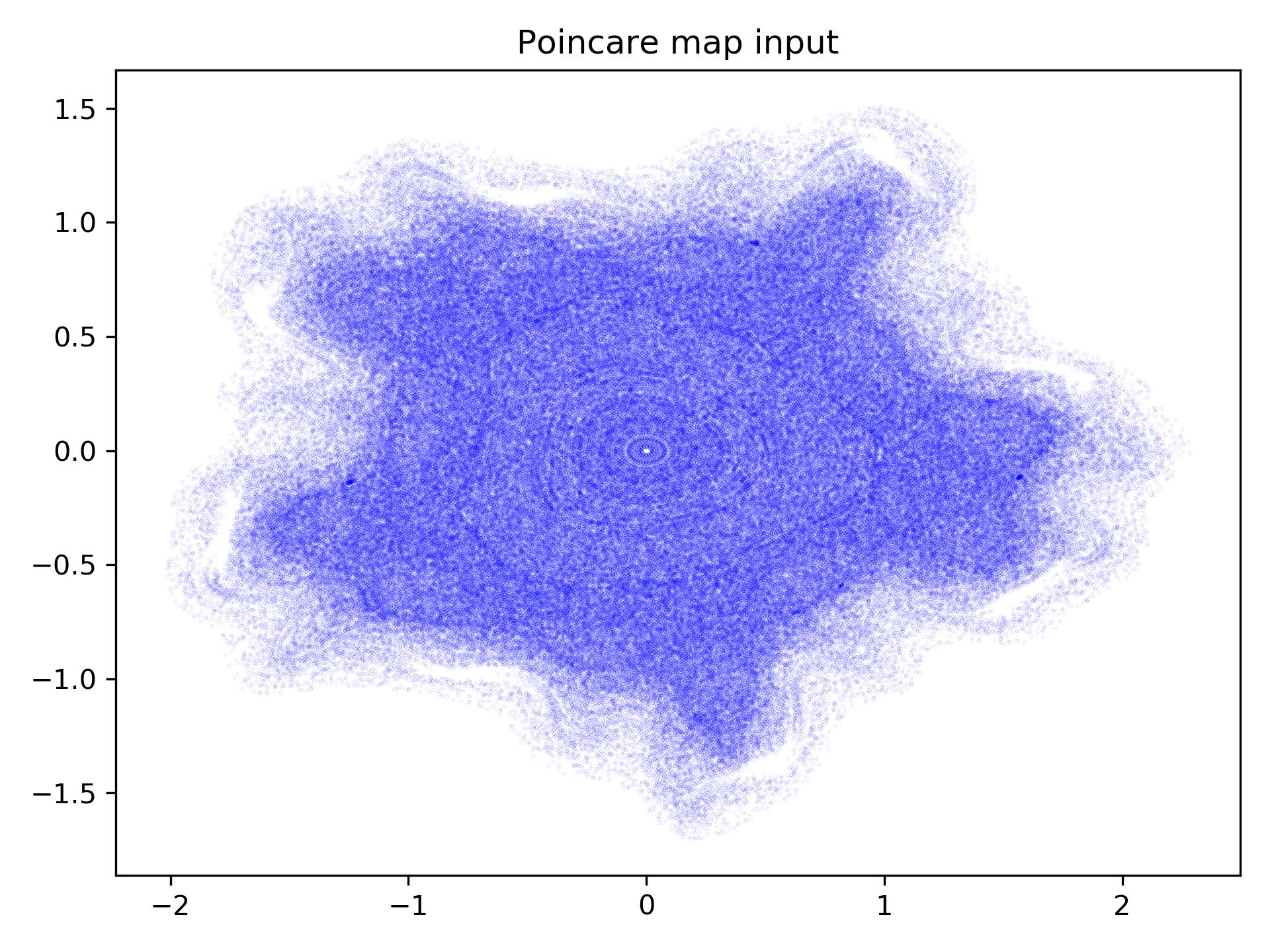} \qquad
\includegraphics[width=.4\textwidth]{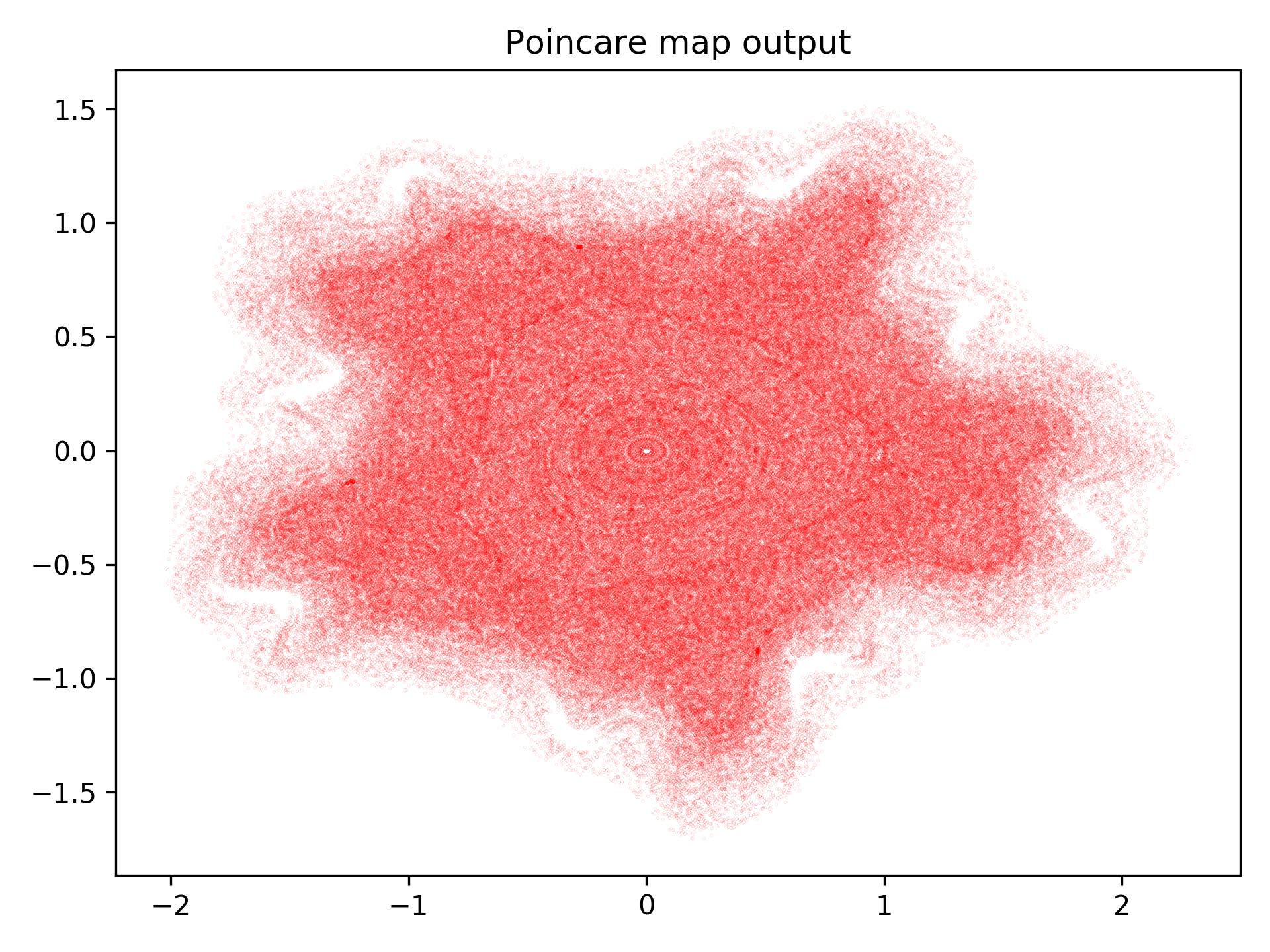}
\end{center}
\caption{Training data in RMP. Left: input of Poincar\'e map. Right: output of Poincar\'e map. A total of 400K points are plotted on each figure.
\label{fig:rmp-data}
}
\end{figure}
}

The \Henon in this case consists of 50 H\'enon layers, each of which has 5 neurons. The network consists of 800 training parameters. 
A deep network is found to be necessary to produce a good Poincar\'e plot in this case, while a less deep network may fail to achieve a loss small enough to produce a comparable Poincar\'e plot.
The number of neurons is reduced  to improve the efficiency of training.
We use the Adam optimizer with 7000 epochs and a decaying learning rate. The initial learning rate is set to be 0.1, and the batch size is 400.
 The training finishes with a final loss of  2.6919e-5. { We evaluated the loss on 4e05 test samples drawn with the same distribution as the test data and found a value of 2.5344e-05}

The trained model is then verified through generating a Poincar\'e plot starting from 20 points along $x$-axis.
The model is used to predict 1000 times recursively, producing the plot presented in Figure~\ref{fig:rmp-inner}. 
We note that two plots match very well with each other, in both of the inner region and chaotic region.
Note that the outer boundaries of the chaotic region match very well, which is in fact challenging to achieve due to the diffusive nature of chaotic dynamics.
It is also interesting to note that there are 13 islands in the chaotic region that the \Henon was able to capture. 
This  inspires us to perform further investigation around those islands.

{
\begin{figure}[htp]
\begin{center}
\includegraphics[width=.32\textwidth]{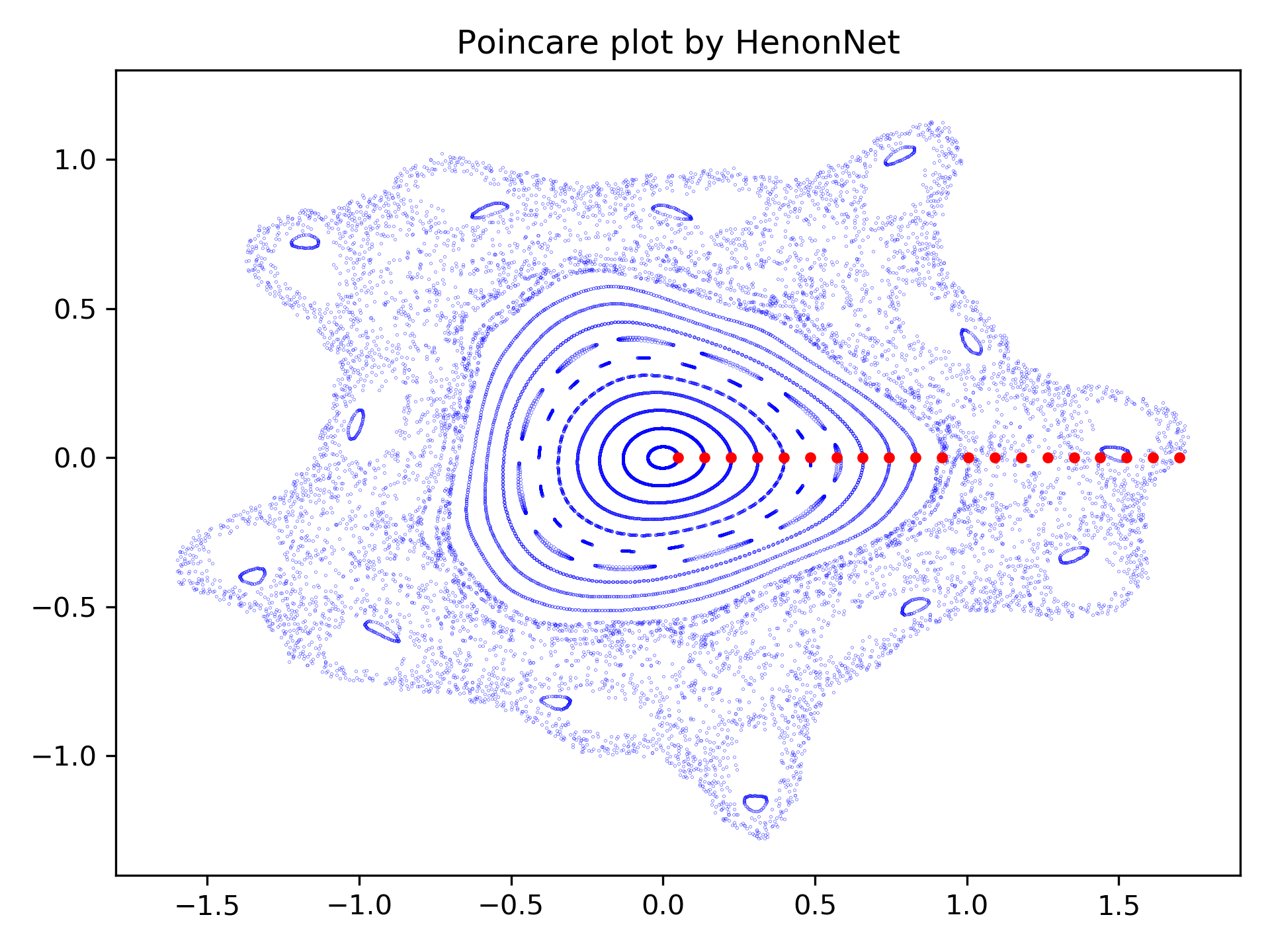}
\includegraphics[width=.32\textwidth]{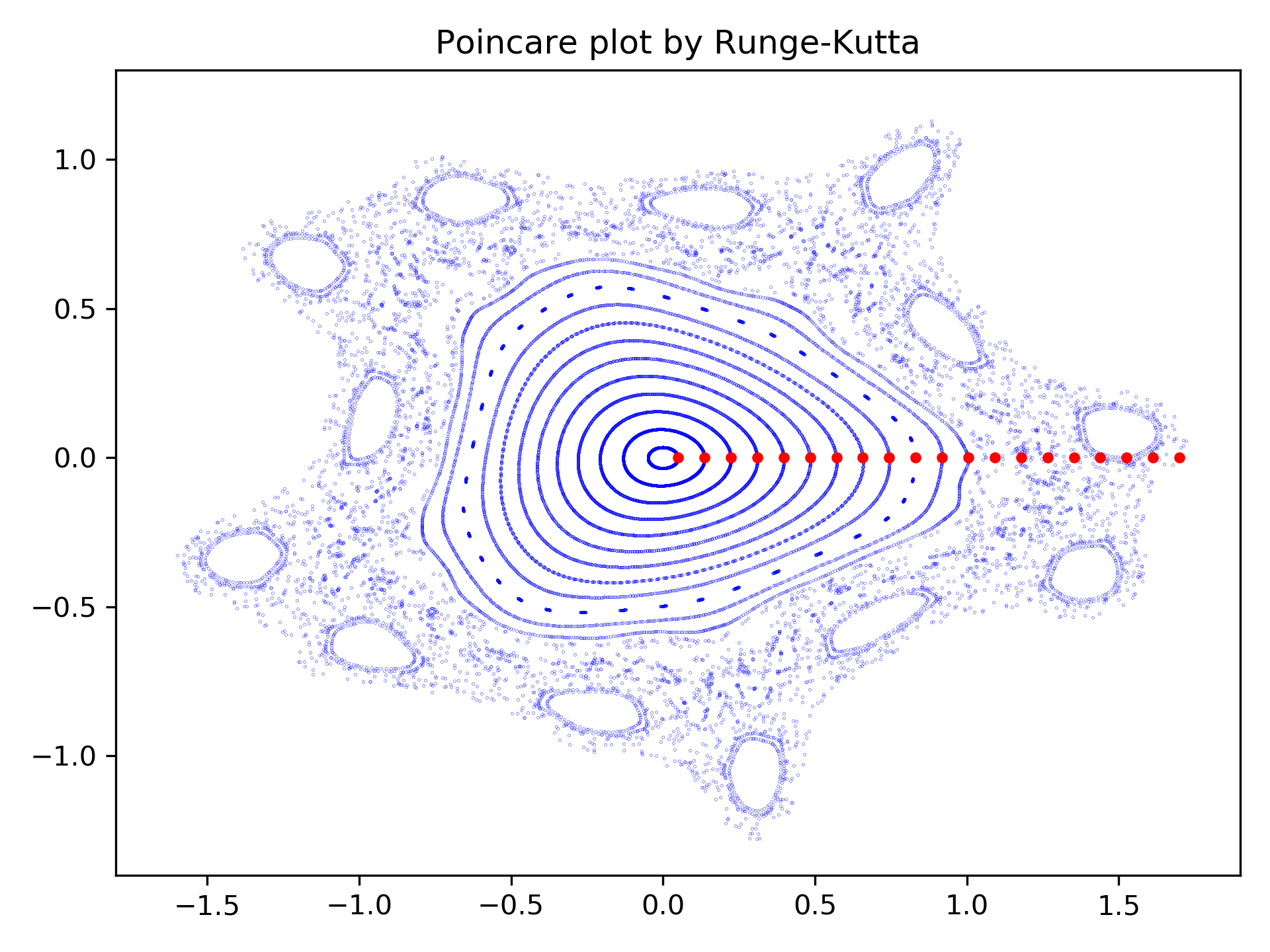}
\includegraphics[width=.32\textwidth]{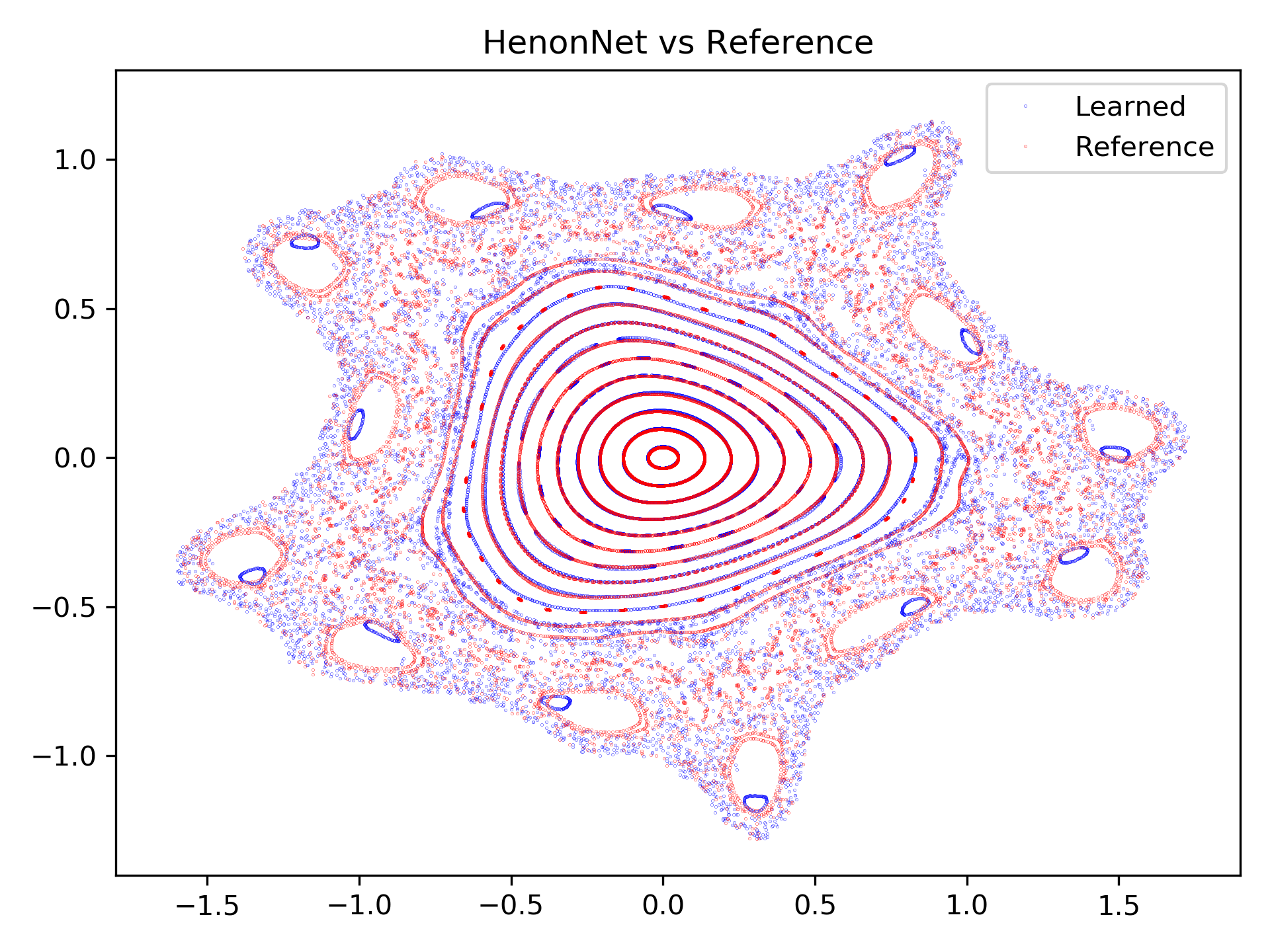}
\end{center}
\caption{Poincar\'e plots  in RMP. Left: Poincar\'e plot generated by the trained H\'enonNet. Middle: Poincar\'e plot generated by RK4. Right: comparing two Poincar\'e plots.
\label{fig:rmp-inner}
}
\end{figure}
}

We shift our attentions to the island chain in the chaotic region.  15 points are selected inside one single island, and then the model is again used to predict 1000 times recursively.
The resulting Poincar\'e plots are presented in Figure~\ref{fig:rmp-islands}.
 We note that the Poincar\'e plot of the \Henon can capture the location of those 13 islands very well, and inside each island 
three smaller islands of comparable size are observed.
On the other hand, in the plot generated by the RK4 method, there are also three smaller islands in a single island but one of them is dominant. 
We notice that the learned internal island dynamics improves with more training. 
 Such a result motivates us to investigate those islands more carefully,
which will be addressed in the next section. 

{
\begin{figure}[htp]
\begin{center}
\includegraphics[width=.45\textwidth]{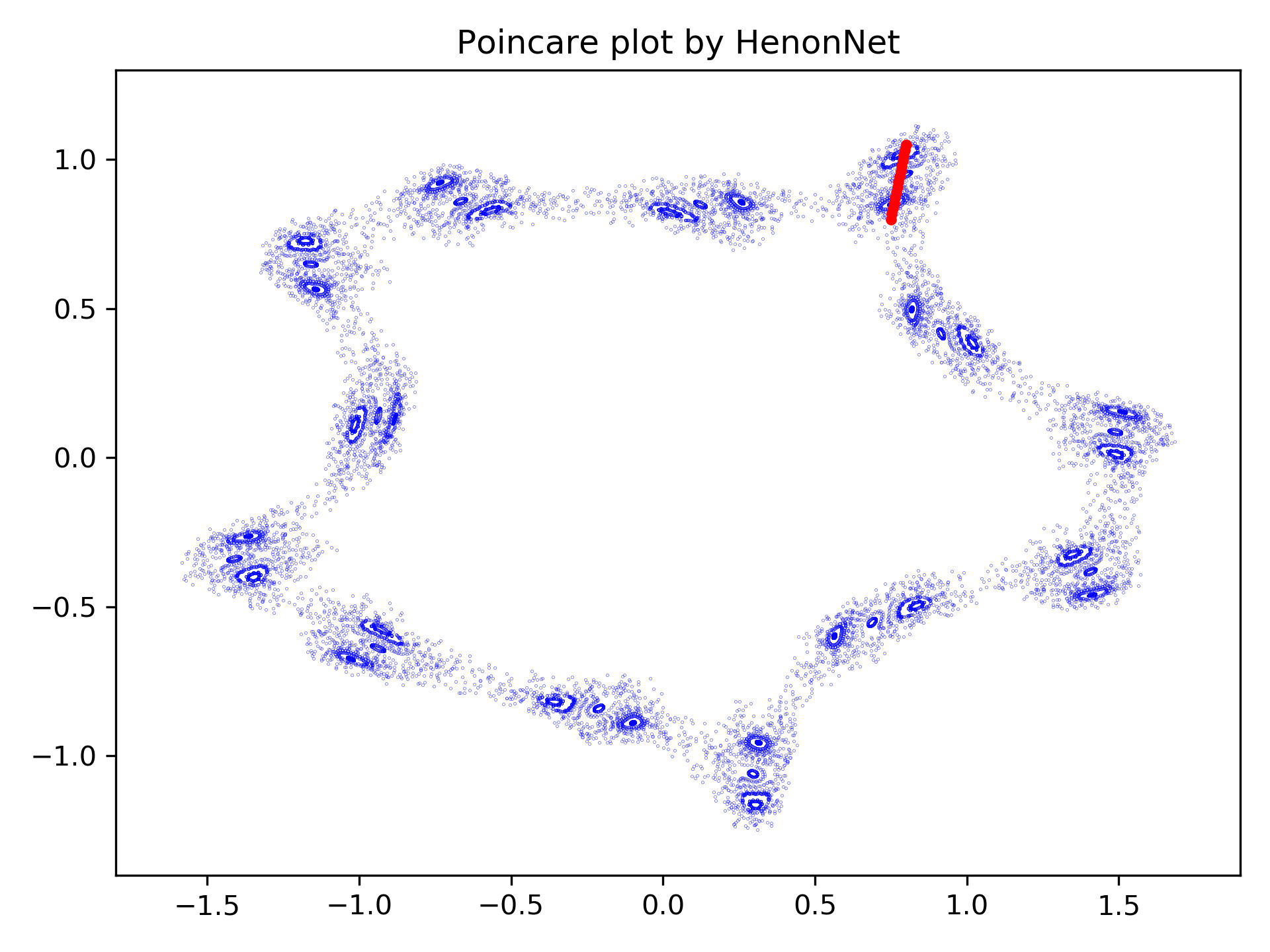} \qquad
\includegraphics[width=.45\textwidth]{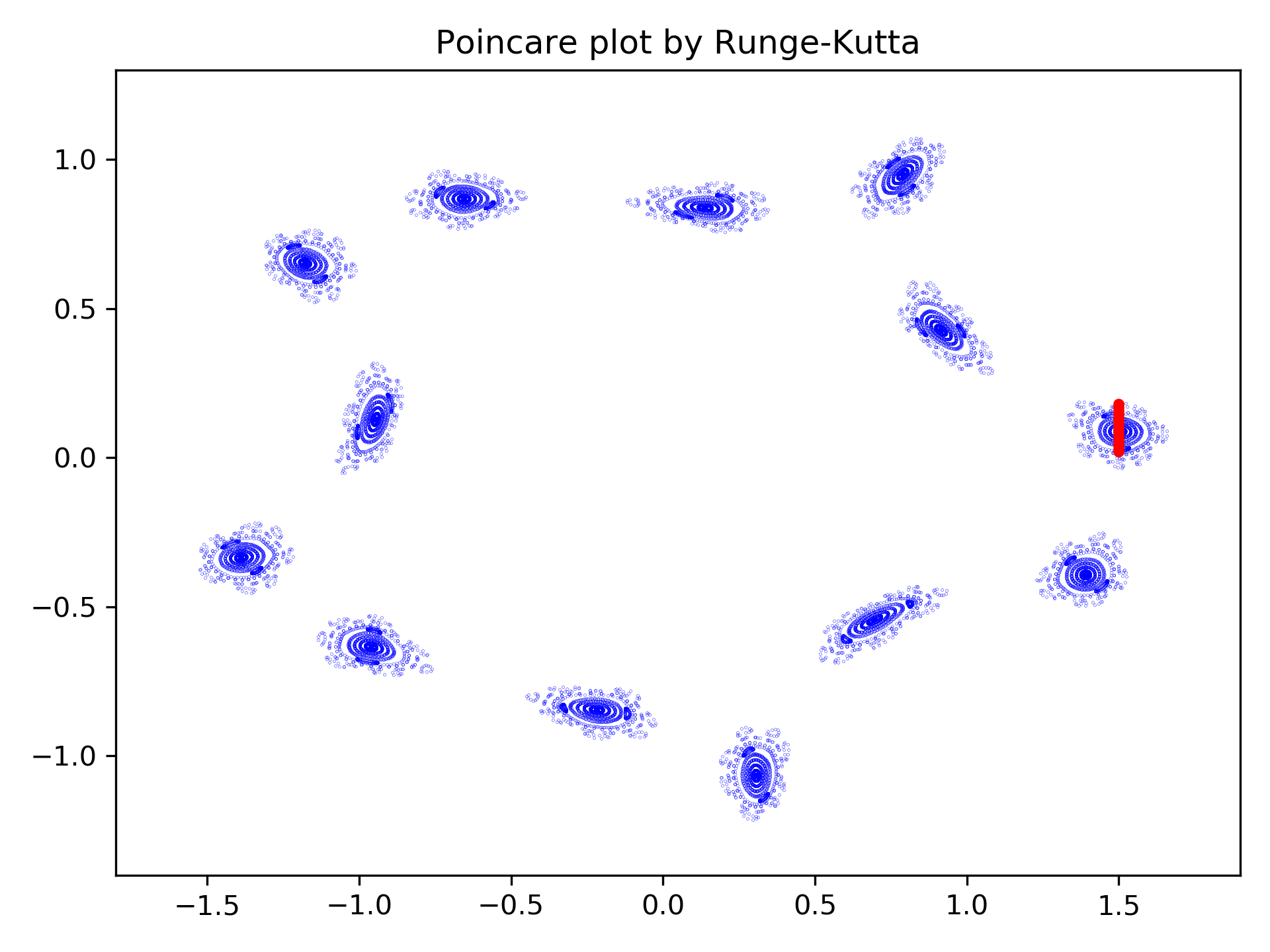}
\end{center}
\caption{Poincar\'e plots of island chain in RMP. Left: Poincar\'e plot generated by the trained H\'enonNet. Right: Poincar\'e plot generated by RK4. 
\label{fig:rmp-islands}
}
\end{figure}
}

 Finally, the \Henon needs 4.26 seconds to predict 1000 times
while the vectorized RK integrator needs 34.2 seconds with 100 time steps for each map. 

\section{A remarkable strategy for creating field-line confinement\label{novel_approx_section}}
\begin{figure}[htp]
\includegraphics[scale = .2]{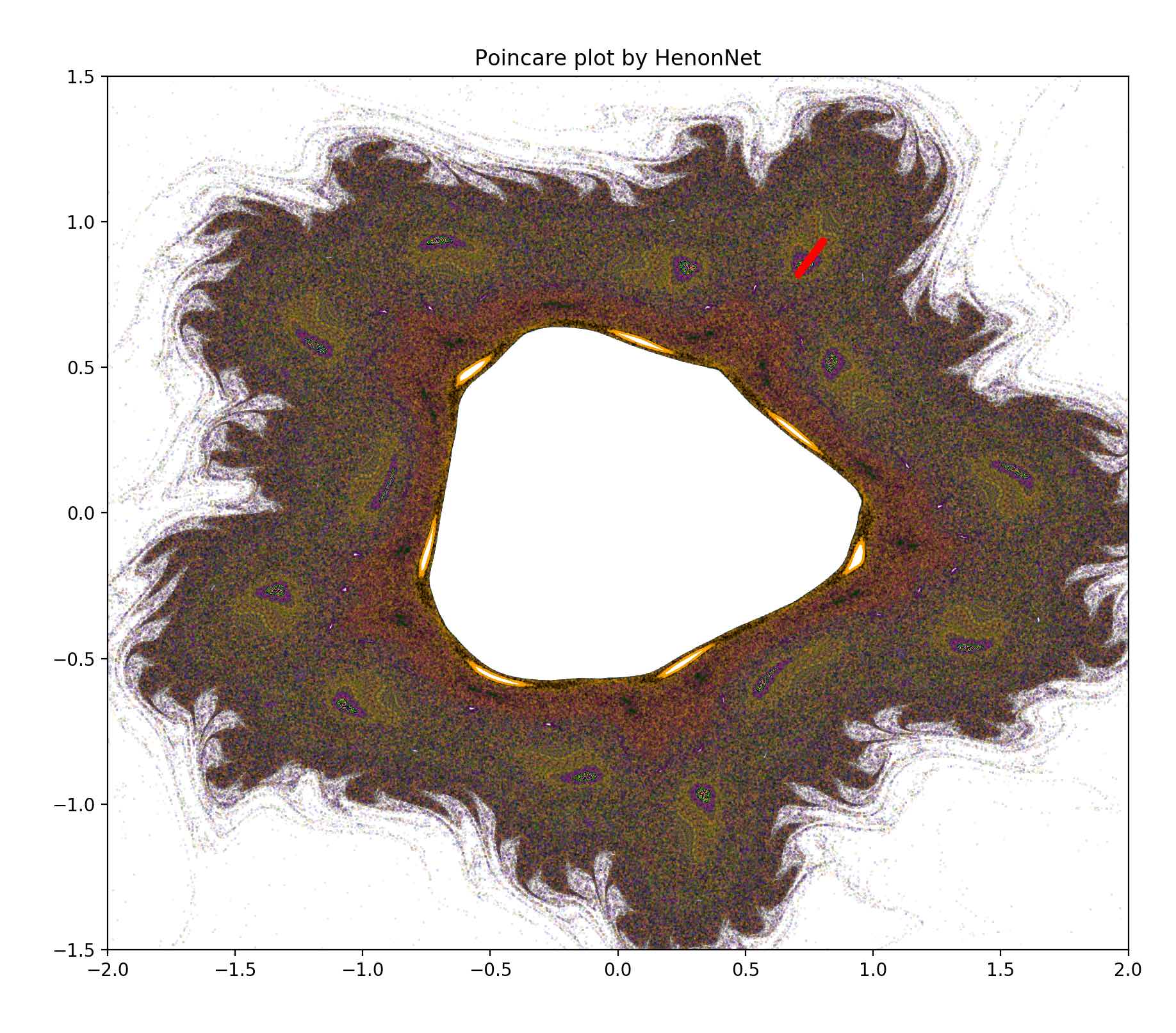}
\caption{A high-resolution Poincar\'e plot generated by a learned Poincar\'e map corresponding to the Hamiltonian defined in Eq.\,\eqref{HRMP}}
\label{chaotic_sea_color}
\end{figure}

In order to gain insight into how a H\'enonNet learns to approximate a Poincar\'e map during the training process, it is interesting to examine the H\'enonNet used to approximate the Poincar\'e map associated with the Hamiltonian \eqref{HRMP} when subjected to 5500 training epochs rather than 7000.  Figure \ref{chaotic_sea_color} shows the corresponding Poincar\'e plot. Initial conditions are indicated by a bright red line. Iterates of a given initial condition share a common marker color. In contrast to the 7000 training-epoch case, the 13-lobe island chain that RK4 predicts should be embedded within the chaotic sea is not clearly delineated by KAM tori.  Instead, variations in the density and color of the plotted markers display impressions of the missing islands. Excesses in marker density, as well as more regular striations of color, indicate that iterations of the learned Poincar\'e map tend to remain near the missing island chain for many iterations of the Poincar\'e map, even though confining KAM tori do not delineate the island boundary.

\begin{figure}[htp]
\includegraphics[scale = .2]{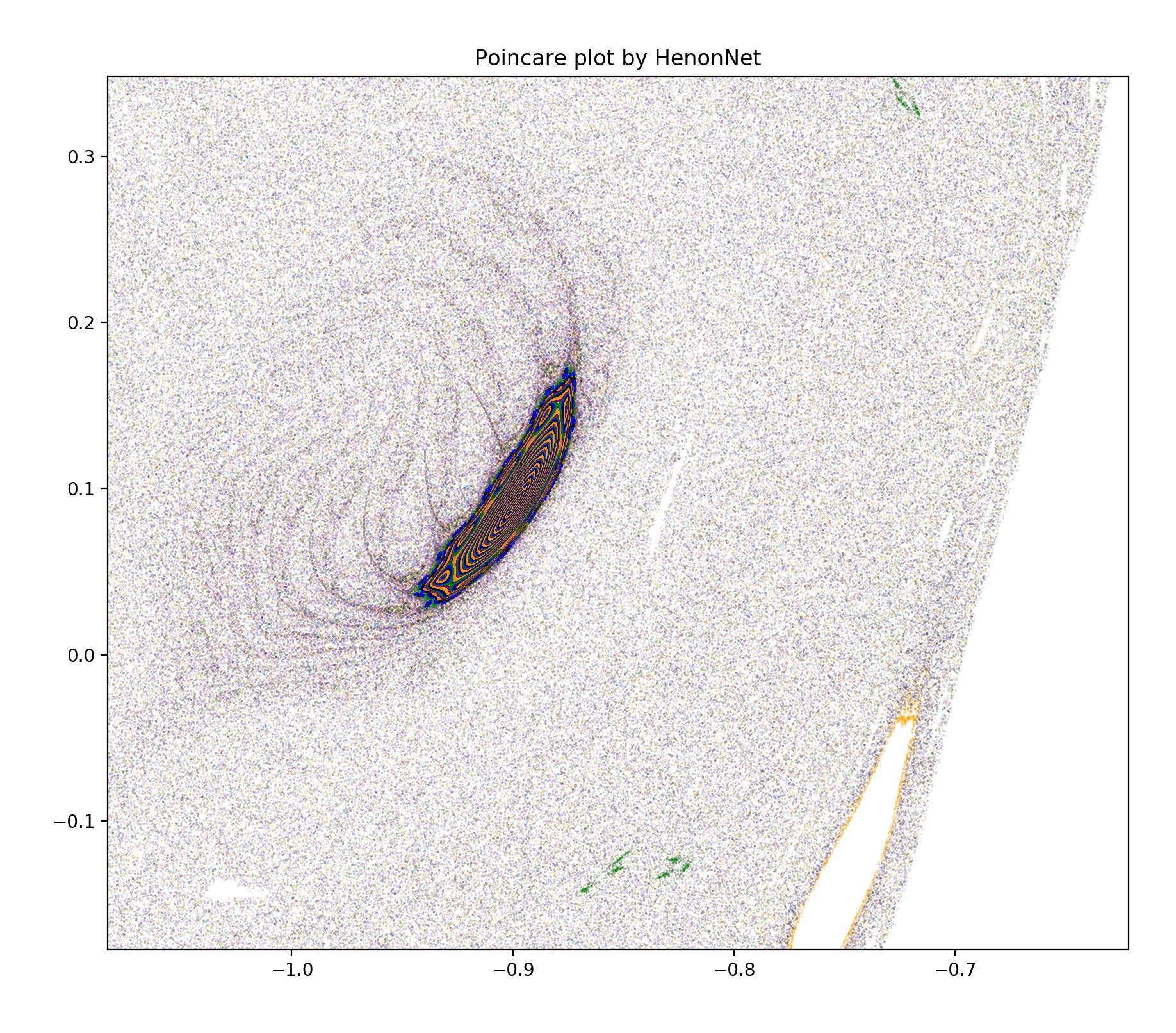}
\caption{Internal structure of the island ``impressions" in Figure \ref{chaotic_sea_color}. The H\'enonNet has learned to mock the confinement properties of a large island chain by coiling hyperbolic invariant manifolds (here outlined by filamentary structures) attached to a much smaller island chain.}
\label{island_zoom_color}
\end{figure}

\begin{figure}[htp]
\includegraphics[scale = .2]{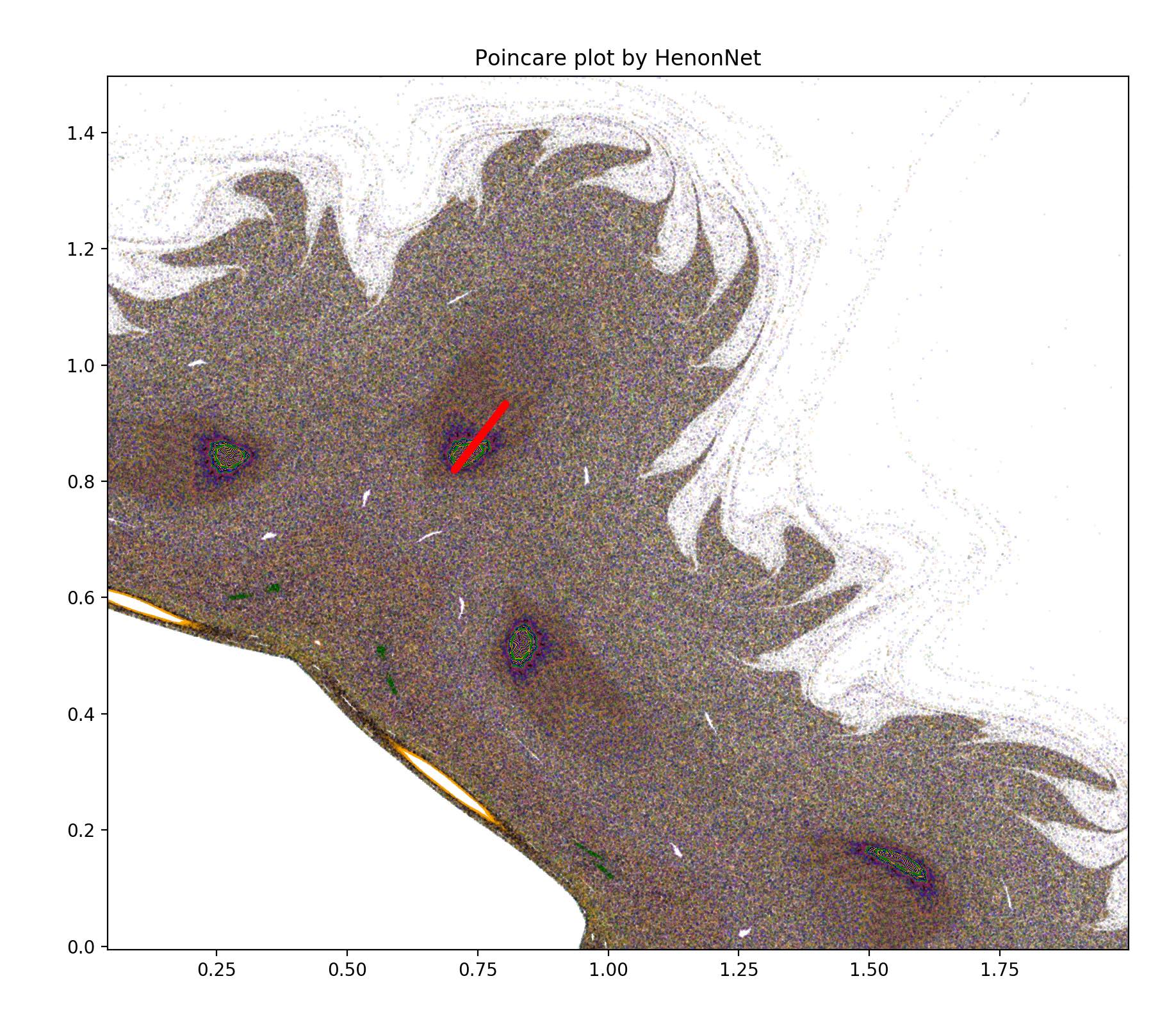}
\caption{A magnified view of Figure \ref{chaotic_sea_color}.}
\label{four_islands_color}
\end{figure}

\begin{figure}
\includegraphics[scale = .2]{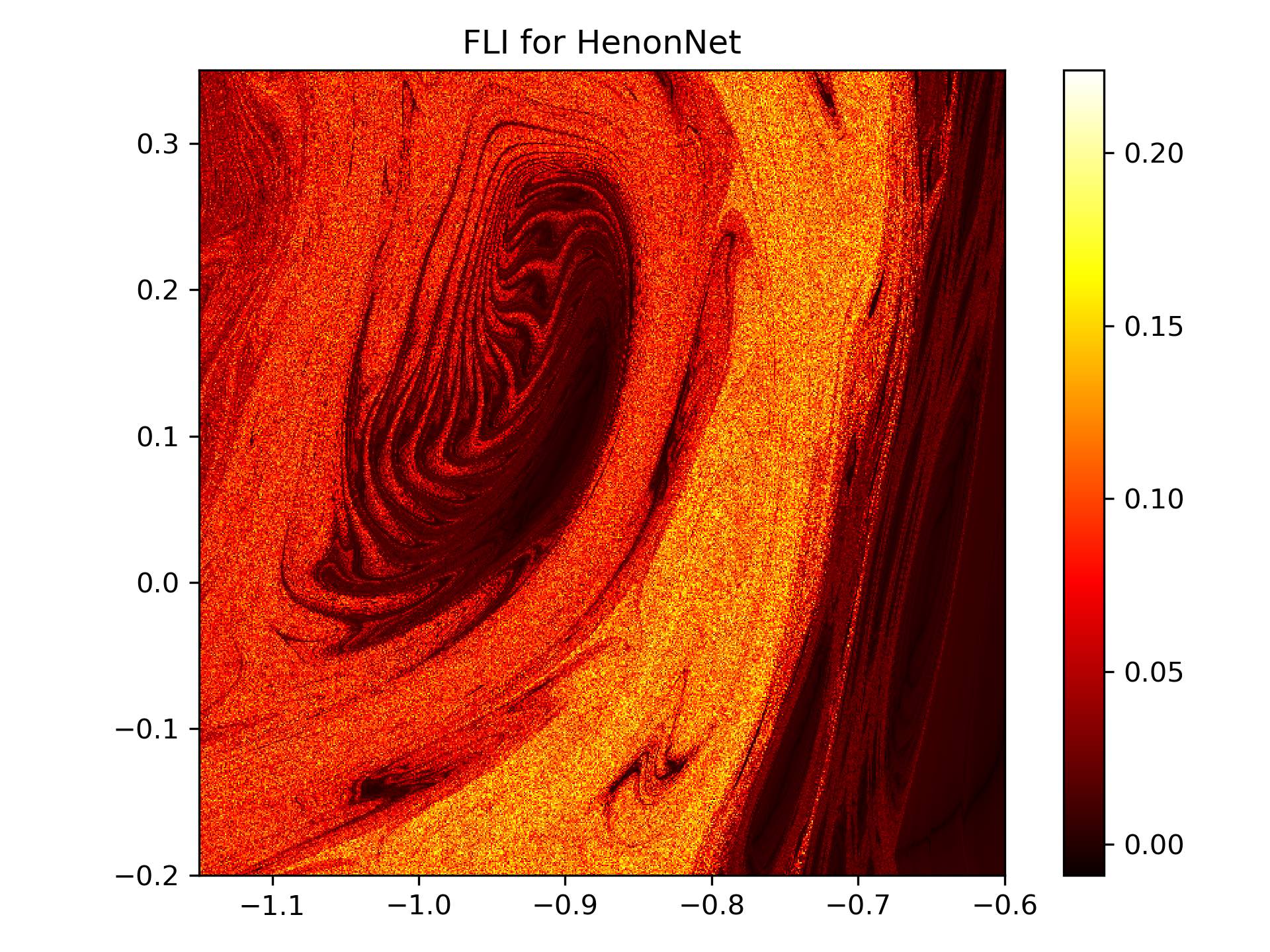}
\caption{Plot of the fast Lyapunov indicator (FLI) in the same region shown in Figure \ref{island_zoom_color}. Darker colors indicate smaller finite-time Lyapunov exponents, and therefore more gradual separation of nearby trajectories. Note that the darkest regions overlap with the filamentary structures depicted in Figure \ref{island_zoom_color}, which is consistent with slower transport in the regions where the filaments are coiled.\label{zoomed_FLI}}
\end{figure}

Figures \ref{island_zoom_color} and \ref{four_islands_color} show magnified views of two regions in the Poinacr\'e plot \ref{chaotic_sea_color} where RK4 integration predicts the presence of embedded islands. Figure \ref{island_zoom_color}, in particular, reveals the detailed internal structure of each island impression. A small subregion of each impression is occupied by a nested family of KAM tori, as indicated by banded orange rings in the figure. Thus, there are in fact conventional islands of stability contained within the larger impressions shown in Figure \ref{chaotic_sea_color}. However, these stability islands only account for a small part of the total impressed area. The bulk of the area occupied by an impression is filled with an array of filamentary structures that emanate from the small region of KAM tori. Each family of these filaments is coiled in such a manner so as to approximately outline the islands predicted by RK4. Presence of these filaments in regions where the rate of chaotic diffusion is depressed suggests that a neighborhood of each filament is ``sticky,'' i.e. that iterations of the learned Poincar\'e map in the neighborhood of a filament will remain near a filament for some time. The sticky-ness of the filamentary structures may also be inferred from Figure \ref{zoomed_FLI}, which shows that the fast Lyapunov indicator (FLI)\cite{Froeschle_2000} is small along the filaments. { The fast Lyapunov indicator provides a finite-time approximation of the largest Lyapunov exponent.} Therefore a natural candidate explanation of the filaments is that they are outlines of stable or unstable manifolds attached to hyperbolic periodic points near the region of KAM tori. We refer the reader to Refs.\,\onlinecite{Meiss_2015} and \onlinecite{MacKay_Meiss_1984} for detailed studies of the phenomenon of partial obstructions to chaotic transport in measure-preserving maps. 

These observations indicate that our H\'enonNet has accomplished a remarkable feat in the course of its training: \emph{it has learned how to imitate the confinement properties of an island chain by coiling hyperbolic invariant manifolds in the region where the island chain should be located.} While hyperbolic invariant sets for area-preserving maps have been studied intensively, and even play a key role in the ``turnstile" physics of divertors in stellarators,\cite{Boozer_Punjabi_2018} to our knowledge the idea of \emph{engineering} the placement of hyperbolic invariant sets in order to create well-confined regions in the field-line phase portrait has never been proposed. 
However, for a H\'enonNet, engineered confinement via coiled hyperbolic manifolds may be easier than achieving a similar level of confinement using KAM tori. Indeed, the H\'enonNet resorted to the coiling strategy \emph{before} (by training epoch 5500)  learning how to approximate the 13-lobe island chain using large KAM tori (which becomes apparent by training epoch 7000). Whatever the mechanism that enabled the H\'enonNet to engineer this (partial) solution to the field-line confinement problem, if distilled and harnessed it could enable a new and more flexible approach {  to the design of magnetic fields with favorable confinement properties.} 
We propose that further study of this mechanism is warranted, and might proceed by careful introspection of the H\'enonNet training process.

\section{Discussion\label{discussion}}

This article demonstrates that Poincar\'e maps for realistic magnetic fields in magnetic fusion devices may be learned by neural networks, and moreover that such learned approximations evaluate orders-of-magnitude faster than traditional field-line following. We advocate a supervised learning strategy, wherein a neural network is shown accurate approximate evaluations of the ground truth Poincar\'e map finely sampled over a given surface-of-section. The novel neural network architecture we have proposed exactly conserves magnetic flux, or, equivalently, exactly satisfies the symplectic condition, and in this sense may be termed ``physics-informed\cite{Raissi_2019}" or ``structure-preserving\cite{Celledoni_arxiv_2020}".  In contrast to the existing symplectic neural network architecture (SympNets) described in Ref.\,\onlinecite{Jin_2020}, our symplectic networks (H\'enonNets) appear to be easier to train, and require less layer depth to achieve a given accuracy. We validated this observation by performing a thorough comparison between the two architectures using a hyperparameter search. Like SympNets, H\'enonNets satisfy a symplectic universal approximation theorem. The approach of using neural networks to directly represent symplectic mappings should be compared with the related approach of learning a system's Hamiltonian\cite{Greydanus_arxiv_2019,Chen_arxiv_2020,Bertalan_2020} or Lagrangian\cite{Qin_arxiv_2019,Cranmer_2020} from data, and then using either a symplectic or variational integrator to construct a corresponding symplectic map. The latter approach will be less able to handle the large timesteps associated with Poincar\'e maps due to stability and accuracy limitations of conventional symplectic and variational integrators.\cite{Morrison_2017,Marsden_2001}

While conducting our numerical experiments using H\'enonNets to approximate Poincar\'e maps, we were lead to test several strategies for making the training process more efficient, involving both the data generation process and the network architecture itself. We found that training is significantly easier when samples of the ground truth Poincar\'e map are drawn from a region that is approximately dynamically invariant. In practice, such sample sets may be generated by applying a few iterations of the ground truth mapping to a given set of sampling points. We also observed that the training process may be simplified by first training with relatively few network layers, and then incrementally increasing the layer depth in subsequent passes of the optimization routine used for training.

In addition, we identified certain strategies for simplifying training that we feel ought to be tested in the future. For example, it may be beneficial to apply a multi-grid-like strategy to the network layers during training. For a H\'enonNet with $2^n$ layers, the first step of layered multi-grid would be to train the $k^{\text{th}}$ layer to map a field line $2\pi/k$ radians around the torus for each $k = 1,\dots,2^n$. Then adjacent layers should be paired (i.e. composed as functions) to form $2^{n-1}$ blocks, and each blocked layer should be trained to map a field line $4\pi/k$ radians around the torus. This blocking process can be continued until there is only a single block to train, after which subsequent subdivision and blocking passes may be applied as needed. Another potentially fruitful avenue for easing the task of learning magnetic field Poincar\'e maps would be to use a H\'enonNet to learn the discrepancy\cite{Kaheman_2019} between the ground truth Poincar\'e map and the so-called tokamap,\cite{Balescu_1998} which is a simple explicit area-preserving map that captures much of the phenomenology of field-line flow observed in toroidal magnetic containment devices. In this approach, the tokamap would be considered a reduced-order model for the magnetic Poincar\'e map, and the H\'enonNet would learn the missing physics. 

{ We remark that the hyperparamers and inner-layer architecture used in H\'enonNets in this work were chosen based on simple heuristics. An extension of a heuristic approach for architecture selection is through automatic neural architecture discovery, \cite{balaprakash2018deephyper} where an outer-loop intelligence may be utilized in evaluating multiple neural network designs based on validation performance. The outer-loop model discovery and evaluation technique is frequently framed as a reinforcement learning or evolutionary algorithm task, where better architectures are gradually discovered at the expense of their poorer counterparts through selection pressure. These strategies are scalable and may be deployed on large clusters and have demonstrated success in outperforming manually design networks for various applications.}

Future analyses of dynamical properties of any given H\'enonNet will be facilitated by the fact that derivatives of a H\'enonNet are readily computed using reverse- or forward-mode automatic differentiation. Thus, important figures of merit, such as Lyapunov exponents, finite-time Lyapunov exponents\cite{Tang_1996}, or fast Lyapunov indicators\cite{Froeschle_2000} (FLI) may be computed with very little algorithmic or computational overhead. In contrast, the computation of derivatives of a Poincar\'e map defined using field-line integration requires significant additions to the field-line integration algorithm, possibly including a solver for the variational equation $\delta\dot{\bm{B}} = \delta\bm{B}\cdot\nabla\bm{B}$. In order to demonstrate this powerful application of neural Poincar\'e maps, we computed the FLI values reported in Figure \ref{zoomed_FLI} using automatic differentiation.

While the primary subject of this article has been teaching networks to learn Poincar\'e maps, we have encountered a phenomenon while training a H\'enonNet to approximate islands embedded in a chaotic sea that may be interpreted as a network teaching its creator new physics. A H\'enonNet learned how to mock the confinement properties of a large island chain by cleverly \emph{coiling} invariant stable and unstable manifolds attached to a much smaller island chain. This proof of principle demonstration, performed by an artificial intelligence, may open the door to new and more flexible methods for designing magnetic fields with good confinement properties. For instance, instead of pursuing confinement by demanding a large volume of KAM tori, one could imagine coiling hyperbolic invariant manifolds in such a manner as to create a ``sticky" region where confinement is desireable.\cite{Contopoulos_2008} However, further work is required to understand the mechanism by which the H\'enonNet accomplished this feat, which may be interpreted as a problem in the area of controlling chaos.\cite{Ciraolo_2004}

\section*{Acknowledgements}
We extend our gratitude to Xianzhu Tang for a number of helpful discussions. This research used resources of the National Energy Research Scientific Computing Center (NERSC), a U.S. Department of Energy Office of Science User Facility. Research presented in this article was supported by the Los Alamos National Laboratory LDRD program under project number 20180756PRD4.
Research was also supported by the U.S. Department of Energy through the Fusion
Theory Program of the Office of Fusion Energy Sciences, and the Tokamak Disruption Simulation (TDS) SciDAC partnerships between the Office of Fusion Energy Science and the Office of Advanced Scientific Computing. We gratefully acknowledge the computing resources provided and operated by the Joint Laboratory for System Evaluation (JLSE) at Argonne National Laboratory. This material is based upon work supported by the U.S. Department of Energy (DOE), Office of Science, Office of Advanced Scientific Computing Research, under Contract DE-AC02-06CH11357.

\section*{Data statement}
The data that support the findings of this study are available from the corresponding author upon reasonable request.

\providecommand{\noopsort}[1]{}\providecommand{\singleletter}[1]{#1}%
%


\end{document}